\documentclass[12pt,a4paper]{article}
\usepackage{amsfonts,latexsym}
\usepackage{graphicx,color}
\usepackage{dcolumn}
\usepackage{graphicx}
\usepackage{amsmath}
\usepackage{amsfonts}
\usepackage{amssymb}
\usepackage{psfrag}
\usepackage{wrapfig}
\usepackage{subfigure}
\usepackage{makeidx}

\renewcommand{\thefootnote}{\fnsymbol{footnote}}

\begin{document}

\vspace{12mm}

\begin{center}
{{{\Large {\bf Charge-dependent scalarization of Einstein- Euler-Heisenberg black holes}}}}\\[10mm]
{Lina Zhang$^1$\footnote{e-mail address: linazhang@hnit.edu.cn},
Yun Soo Myung$^2$\footnote{e-mail address: ysmyung@inje.ac.kr},
De-Cheng Zou$^3$\footnote{e-mail address: dczou@jxnu.edu.cn}, Chao-Ming Zhang$^{4}$\footnote{e-mail address: chaomingzhang70@gmail.com}, 
Sheng-Yuan Li$^{5}$\footnote{shengyuanli77@outlook.com}} 

{${}^1$College of Science, Hunan Institute of Technology, Hengyang 421002, China\\[0pt]}

{${}^2$Center for Quantum Spacetime, Sogang University, Seoul 04107, Republic of  Korea\\[0pt]}

{${}^3$School of Physics, Jiangxi Normal University, Nanchang 330022, China\\[0pt]}

{${}^4${Center for Relativistic Astrophysics and High Energy Physics, Nanchang University, Nanchang 330031, China\\[0pt] }
{$^{5}$Center for Gravitation and Cosmology, College of Physical Science and Technology, Yangzhou University, Yangzhou 225009, China}}
\end{center}
\vspace{2mm}

\begin{abstract}
Charge-dependent scalarization of the  Einstein-Euler-Heisenberg (EEH) black hole is carried out  in the EEH-scalar theory by introducing an exponential scalar coupling with coupling constant $\alpha$ to the Maxwell and nonlinear electrodynamic terms.  The bald black hole (EEHBH)  is  described by mass $M$ and  arbitrary  magnetic charge $q$ and has a single horizon  when choosing the action parameter $\mu=0.3$. {\bf  Here, we find the critical onset charge $q_c=r_+^2/\sqrt{2\mu}$ with $r_+(M,q_c)$ the outer horizon radius from the resonance condition.}
The spontaneous scalarization ($\alpha^+$) of this black hole is available  for  charge $0<q< q_c=1.115$ {\bf for $M=1$ and $\mu=0.3$}, with positive $\alpha$, whereas  its new scalarization  ($\alpha^-$) occurs for $q> q_c$ and negative $\alpha$.  The former case of $q=0.5$  implies infinite branches of scalarized EEHBHs but its fundamental branch ($n=0$) is stable against the $l=0$ modes considered here, while for $q=2$ and $q=20$, one stable scalarized branch is found
for each charge within the explored parameter ranges.

\end{abstract}

\vspace{1.5cm}

\hspace{11.5cm}
\renewcommand{\thefootnote}{\arabic{footnote}}
\setcounter{footnote}{0}

\section{Introduction}
Euler and Heisenberg (EH) have  proposed a novel framework in which one-loop corrections are incorporated into quantum electrodynamics (QED) with  the EH parameter $a=8\alpha_{fs}/45m^4(\equiv8\mu)$ to explain the vacuum polarization in QED and control the strength of the term nonlinear electrodynamics (NED: $\mathcal{F}^2$)~\cite{Heisenberg:1936nmg}. 
Yajima and Tamaki have obtained  the Einstein-Euler-Heisenberg (EEH) black hole solution described by mass ($M$), charge ($q$), and the EH parameter ($\mu$) when considering Einstein-Maxwell theory with the NED term (EEH theory)~\cite{Yajima:2000kw}.   One feature of this solution states that  the black hole charge can naturally extend to accommodate  the regime of $q \geq M$ with an appropriate choice of $\mu$. In the context of the EEH  theory, it is interesting to  note   the potential detection of quantum gravity effects~\cite{Brodin:2001zz,Allahyari:2019jqz,Kruglov:2020tes}.  Furthermore, the general NED-black solution was obtained from $\mathcal{F}^n$ term~\cite{Lee:2026cit}. 

On the other hand, no-hair theorem states that a black hole can be  described by mass ($M$), electric charge ($Q$), and rotation parameter ($a$)~\cite{Ruffini:1971bza}.
If a scalar field is minimally coupled to  gravitational and electromagnetic  fields, the scalar could not survive  as an equilibrium configuration around the black hole, describing   no-scalar hair theorem~\cite{Herdeiro:2015waa}.
However, by introducing a conformal (nonminimal) scalar coupling
to the Ricci scalar, one obtains the extremal BBMB black hole
with secondary scalar hair~\cite{Bocharova:1970skc,Bekenstein:1974sf}.
Importantly, spontaneous scalarization for a nonminimal scalar coupling to Gauss-Bonnet term~\cite{Doneva:2017bvd,Silva:2017uqg,Antoniou:2017acq} or Maxwell term~\cite{Herdeiro:2018wub} were realized through  triggering   by tachyonic scalar.

Scalarization provides a dynamical mechanism for the formation of hairy black holes and may play   a key role  to understand the interaction between gravity and matter. It was mainly determined by potentials for the minimal coupling to gravity and  forms of the coupling function to  matter for nonminimal couplings. 
Concerning  scalarizations of EEH black holes, it is worth noting that a charged hairy black hole was obtained analytically  from the EEH-scalar (EEHS) theory when considering  a complicated scalar  potential~\cite{Karakasis:2022xzm}.  We note that the scalarization of Reissner-Norstr\"om black hole (RNBH) was studied in the EEHS theory with a quadratic scalar coupling $f(\phi)=\beta^2 \phi^2$ to the other Maxwell  and EH terms~\cite{Kiorpelidi:2023jjw}.  {\bf In this case,  the RNBH  is regarded as the bald solution with $\phi=0$. The scalarization mechanism is closely analogous and  the main
new feature of present work is the EEH bald background and the resulting charge-dependent change between positive- and negative-coupling scalarization.} 

Recently, a negative potential-induced scalarization of the EEH black hole with single horizon ($\mu=0.3$) was investigated  with unlimited charge $q$ in the EEHS theory~\cite{Guo:2025ksj}. For this purpose, a negative potential of $V(\phi) = -\alpha^2 \phi^6$  was  introduced. It was known to be one of the simplest forms which  can obtain scalarized black holes, even in the Einstein–minimally coupled scalar theory \cite{Chew:2024evh}. It turned out that the single branch of scalarized black holes was unstable against  radial perturbations when  computing quasinormal frequencies of a perturbed scalar. Curiously, however,  one observed that for a choice of  mass $M=1/2$, the scalar charge $q_s$ exhibits a primary hair for $ q < 1/2$, whereas  it becomes a constant (secondary hair) for $q>1/2$.
Furthermore, spontaneous scalarizations of the  EEH  black hole with single horizon  were  performed for $q=0.5,~2,~20$ in the EEHS theory by introducing an exponential coupling of $e^{-\alpha \phi^2}$  to the Maxwell term only~\cite{Zhang:2025msi}.  
It was found that for $M=1$, there is some difference in the onset of scalarization between $q\le1$ and $q>1$. 
In this case, infinite branches labeled by the number of $n=0,1,2,\cdots$  of  scalarized EEH black holes were  allowed  by taking into account  infinite scalar clouds  appeared around  the EEH black hole.
The  fundamental branch ($n=0$) of scalarized  EEH black holes  is stable against the $l=0$ modes considered here, whereas one excited  branch ($n=1$) is unstable. It is worth noting that there was no significant  difference in spontaneous scalarization among $q=0.5,~2,~20$ cases.

The previous two studies ask us a central question on  what  $q$-dependent scalarization of EEH black hole with single horizon looks really like. 

In the present work,  we wish  to investigate the charge ($q$)-dependent scalarization of the EEH  black holes with single horizon  within  the EEHS theory by introducing an exponential scalar coupling to the Maxwell and NED terms.   The bald black hole (EEHBH)  is still  described by mass $M$ and  unrestricted   magnetic charge $q$ and it possesses the  single horizon  for a choice of the  action parameter $\mu=0.3$.  
Importantly,  we find that infinite branches of scalarized EEHBHs are  allowed  for a range of  charge $0<q< q_c$ with critical onset charge $q_c=1.115$ and  $\alpha>0$ through spontaneous scalarization ($\alpha^+$), whereas  a new  scalarization  ($\alpha^-$) might  occur for $q>q_c$ and  $\alpha<0$. For the representative cases $q=2$ and $q=20$, one scalarized branch is found for each charge within the explored parameter ranges. The fundamental branch ($n=0$) of the former is stable against the $l=0$ modes considered here, while for the latter cases, the scalarized branches obtained for $q=2$ and $q=20$ are stable against the $l=0$ modes considered here. This will clarify the charge-dependent scalarization of EEHBHs with single horizon by considering {\bf single  coupling to Maxwell and NED terms.}

\section{EEHS theory and its onset scalarizations}

We start with  the Einstein-Euler-Heisenberg-scalar (EEHS) theory with an action parameter $\mu$ to the  NED term ($\mathcal{F}^2$)
\begin{equation}
S_{\rm EEHS}=\frac{1}{16 \pi}\int d^4 x\sqrt{-g}\Big[ R-2\partial_\mu \phi \partial^\mu \phi-e^{-\alpha \phi^2} (\mathcal{F}-\mu \mathcal{F}^2)\Big],\label{Act1}
\end{equation}
where $\alpha$ is a scalar coupling constant to the Maxwell ($\mathcal{F}=F_{\mu\nu} F^{\mu\nu}$) and NED terms. {\bf At this stage, we emphasize that   the NED term ($\mathcal{F}^2$)  is treated as part of a phenomenological NED model rather than as a perturbative truncation of the Euler–Heisenberg effective action.}
The bald black hole solutions were discussed in the  EEH theory without scalar~\cite{Yajima:2000kw,Allahyari:2019jqz,Amaro:2020xro,Breton:2021mju}.
The Einstein  equation is derived from the action (\ref{Act1})
\begin{eqnarray}
 G_{\mu\nu}=T^\phi_{\mu\nu}+T_{\mu\nu}^{F} \label{equa1}
\end{eqnarray}
with its energy-momentum tensors
\begin{eqnarray}
&&T_{\mu\nu}^\phi=2\Big[\partial _\mu \phi\partial _\nu \phi -\frac{1}{2}(\partial \phi)^2g_{\mu\nu}\Big], \nonumber \\
&&T_{\mu\nu}^F=2e^{-\alpha \phi^2}\Big[F_{\mu\rho}F_{\nu}~^\rho -2\mu \mathcal{F}F_{\mu\rho}F^\rho_\nu-\frac{1}{4}\mathcal{F}(1-\mu \mathcal{F})g_{\mu\nu}\Big] \label{emten}.
\end{eqnarray}
The Maxwell equation is 
\begin{eqnarray} \label{M-eq}
&&\nabla_\mu (F^{\mu\nu}-2\mu\mathcal{F}F^{\mu\nu})=2\alpha \phi\nabla_{\mu} (\phi)F^{\mu\nu}(1-2\mu \mathcal{F}).\label{M-eq1}
\end{eqnarray}
Importantly, the scalar equation is given by
\begin{equation}
\square \phi +\frac{\alpha}{2}\mathcal{F}(1-\mu \mathcal{F})e^{-\alpha \phi^2}\phi=0 \label{s-equa}.
\end{equation}
Considering the mass function $\bar{m}(r)$ together with $\bar{A}_{\hat{\varphi}}=-q\cos\theta ~(\bar{\mathcal{F}}=\frac{2q^2}{r^4})$ and $\bar{\phi}=0$, the $({}_t~^t)$-component of the Einstein equation leads to
\begin{equation}
\bar{m}'(r)=\frac{q^2}{2r^2}-\mu \frac{q^4}{r^6}. \label{mass-eq}
\end{equation}
Solving the above equation leads to the bald black hole  (EEHBH) solution
\begin{eqnarray}
ds^2_{\rm EEHBH}=\bar{g}_{\mu\nu}dx^\mu dx^\nu=-f(r) dt^2+\frac{dr^2}{f(r)} +r^2d\Omega^2_2   \label{EEH-s}
\end{eqnarray}
with
\begin{equation} \label{metric-func}
 f(r)\equiv1-\frac{2\bar{m}(r,q,\mu)}{r}=1-\frac{2M}{r}+\frac{q^2}{r^2}-\frac{2\mu q^4}{5r^6}.
\end{equation}
 This black hole  solution is  described by three parameters $(M,q,\mu)$
where $M$  denotes the ADM mass, $q$ is the magnetic  charge, and $\mu$ is the action parameter to represent  the strength of  NED term.
 Hereafter, we choose  $\mu= 0.3$ to find a black hole  with single horizon.  In case of $\mu\le 0.08$ with $M=1$, however, there exist multiple  horizons. This makes scalarization analysis  difficult~\cite{Myung:2025zxu}.
 Here, it is important to note that  there is no constraint on the magnetic charge $q$ and thus, its single root $r_{+}(M=1,q)$ obtained from  $f(r)=0$ becomes a continuous  function of $q$, which differs quite  from the RNBH for  $\mu=0$ with two (outer/inner) horizons $r_{\rm RN\pm}(M=1,q)=1\pm\sqrt{1-q^2}$ [see (Left) Fig. 1]. 
\begin{figure*}[t!]
   \centering
  \includegraphics[width=0.4\textwidth]{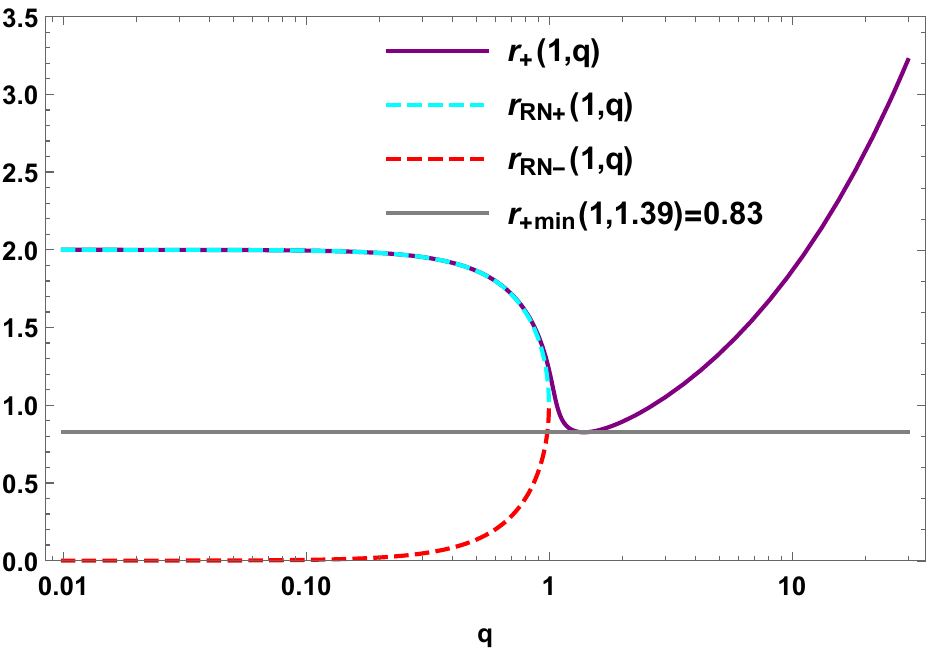}
   \hfill%
\includegraphics[width=0.4\textwidth]{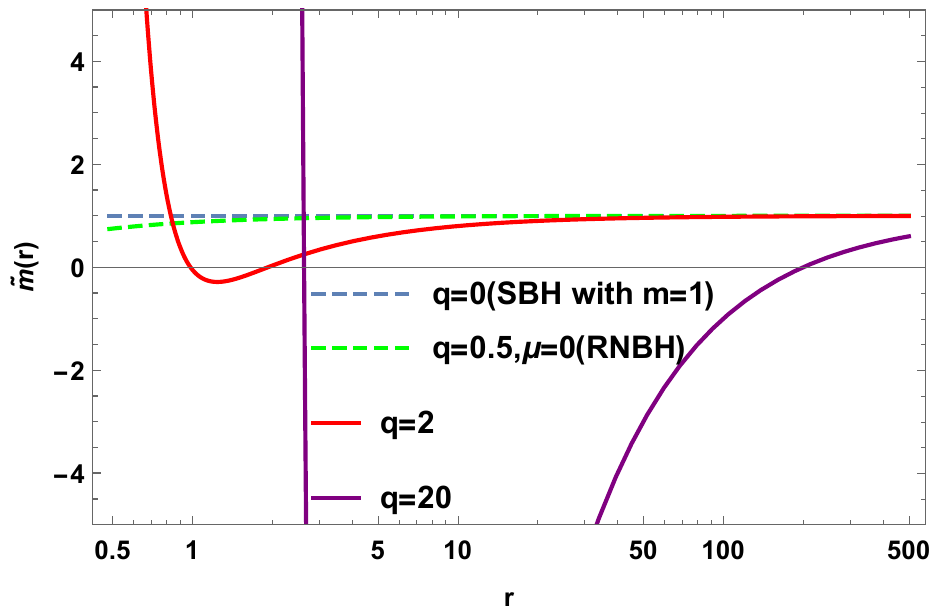}
\caption{(Left) Outer horizons $r_+(M=1,q),~r_{\rm RN+}(1,q\in[0,1])$ and  an RN inner horizon $r_{\rm RN-}(1,q\in[0,1])$. Here, $r_+(1,q)$ takes  the minimum value $r_+(1,1.39)=0.83$ and then, it is an increasing function of $q$.
(Right) Four  mass functions with $m=1$ for SBH with $q=0$: $\tilde{m}(r,q=0.5,\mu=0)\simeq \tilde{m}(r,q=0.5,\mu=0.3) $ for RNBH,  $\tilde{m}(r,q=2,\mu=0.3)$, and $\tilde{m}(r,q=20,\mu=0.3)$ for EEHBH.}   
\end{figure*}

Also, we note that  its thermodynamics with  $\mu= 0.3$ is quite different from that of the RNBH, implying no Davies point in heat capacity. In addition,  the mass function contains all information to form the EEHBH.  (Right) Fig. 1 shows that the mass function $\tilde{m}(r,q,\mu)$ takes  different shapes for different $q$. For $q>1(=2,20)$ with $\mu=0.3$, they have four zero crossing points, differing from $q=0$ (Schwarzschild BH: SBH) and $q=0.5,\mu=0$ (RNBH).

Let us consider  perturbations around the EEHBH background
\begin{equation}
g_{\mu\nu}=\bar{g}_{\mu\nu}+h_{\mu\nu},\quad\phi=0+\delta \varphi, \quad F_{\mu\nu}=\bar{F}_{\mu\nu}+f_{\mu\nu},\quad f_{\mu\nu} =\partial_\mu a_\nu-\partial_\nu a_\mu.
\end{equation}
Before we proceed, we note that the linearized EEH theory leads to being stable against the metric-vector perturbations for an electrically charged EEHBH~\cite{Luo:2022gdz}.  Here, its metric function can be  obtained when replacing  $\mu$ and $q$ by $8\mu$ and $Q_e$.   Hence, it would be better to focus on solving  the  linearized scalar equation which determines the tachyonic instability of the scalar around EEHBH
\begin{equation}
[\bar{\square}-m^2_{\rm eff}]\delta \varphi=0,\quad m^2_{\rm eff}(r,q)=-\alpha \Big(\frac{q^2}{r^4}-\frac{2\mu q^4}{r^8}\Big).\label{per-eq}
\end{equation}
\begin{figure*}[t!]
   \centering
  \includegraphics[width=0.4\textwidth]{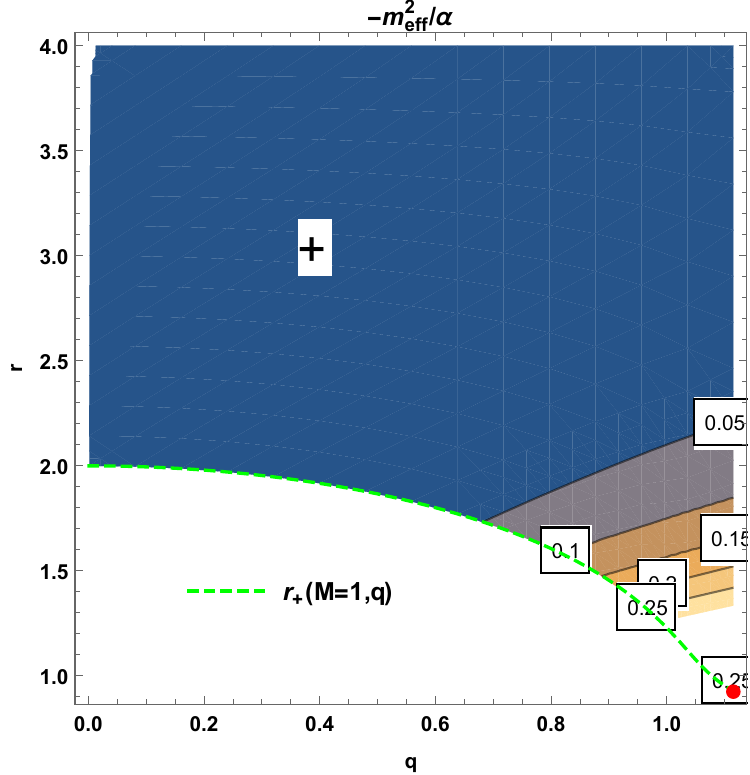}
   \hfill%
\includegraphics[width=0.4\textwidth]{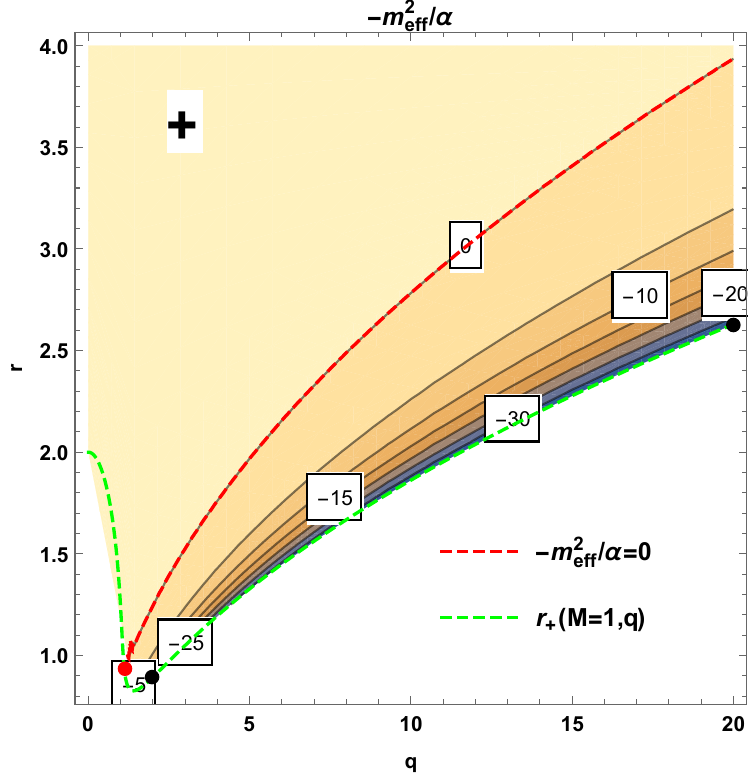}
\caption{ (Left) $- m^2_{\rm eff}(r,q)/\alpha$  as functions of
$r\in [r_+(M=1,q),4]$ and $q\in[0,1.115]$ but its zero  is not allowed. One finds that $m^2_{\rm eff}<0$ for $0<q< 1.115$ and $\alpha>0$. A red dot denotes the end point of  $[q_c,r_+(1,q_c)]$. (Right)  $- m^2_{\rm eff}(r,q)/\alpha$  as functions of
$r\in [r_+(M=1,q),4]$ and $q\in[1.115,20]$ and its zero (red curve) is available, starting from a red dot at $[q_c,r_+(1,q_c)]$. Clearly, it is shown that $m^2_{\rm eff}<0$ is achieved for $q>1.115$ and $\alpha<0$. }
\end{figure*}
As shown in Fig. 2, one finds that  $m^2_{\rm eff}(r,q)<0$ can be achieved in the near-horizon  for $0<q\le q_c(= 1.115)$ and $\alpha>0$, while $m^2_{\rm eff}(r,q)<0$ is achieved for $q> q_c$ and $\alpha<0$.
Here, $q_c$ represents an important charge that satisfies $- m^2_{\rm eff}(r_+(1,q_c),q_c)/\alpha=0$ and it will be identified with  a critical onset charge.
It is desirable to note a red point and two black points: $- m^2_{\rm eff}(r_+(M=1,q),q)/\alpha=0(q=q_c),~-17.34(q=2),~-33.64(q=20)$, which are located  on the outer horizon.

We introduce a tortoise coordinate defined by $dr_*=dr/f(r)$ and consider a separation of variables
\begin{equation}
\delta\phi(t,r_*,\theta,\varphi)=\sum_m\sum^\infty_{l=|m|}\frac{\varphi_{lm}(t,r_*)}{r}Y_{lm}(\theta,\varphi).
\end{equation}
Its $s(l=0,m=0)$-mode linearized equation reduces  to
\begin{equation} \label{mode-d}
\frac{\partial^2\varphi_{00}(t,r_*)}{\partial r_*^2} -\frac{\partial^2\varphi_{00}(t,r_*)}{\partial t^2}=V_{\rm EEH}(r)\varphi_{00}(t,r_*),
\end{equation}
whose  potential is given by
\begin{eqnarray}
V_{\rm EEH}(r,M,q,\alpha)&=&f(r)\Big[\frac{2M}{r^3}-\frac{2q^2}{r^4}+\frac{12\mu q^4}{5r^8}+m^2_{\rm eff}\Big].\label{EEH-P}
\end{eqnarray}
For $0<q<q_c$ and  $\varphi_{00}(t,r_*)\sim u(r_*)e^{-i\omega t}$, $\alpha^+$ scalarization of EEHBH is allowed   for positive $\alpha$.  See the case for $\mu=0$ in ~\cite{Herdeiro:2018wub,Myung:2018vug,Myung:2018jvi}.
On the other hand,  one expects to find  $\alpha^-$ scalarization for $q>q_c$ and negative $\alpha$, which is very similar to GB$^-$ scalarization~\cite{Brihaye:2019kvj,Herdeiro:2021vjo}.

Before we proceed, we are in a  position   to find the  critical onset charge $q_c$, which determines the lower bound ($q> q_c$) for the  onset  $\alpha^-$ scalarization  by making use of  the Hod's approach~\cite{Hod:2020jjy}.
To obtain the critical onset parameter, it is enough to consider the potential term:  $V_{\rm EEH}(r)$ $\cdot \varphi_{00}(t,r_*)=0$.
The critical onset point   is defined actually  by the critical black hole which  denotes   the boundary between EEHBH  and  scalarized EEHBH  in the limit of $\alpha \to -\infty$.
In this limit, it is represented  by a degenerate  binding potential well  whose two turning points
merge at the outer horizon as
\begin{eqnarray}
rc(M,q)\cdot \varphi_{00}(t,r_*)=0,\quad rc(M,q)= \frac{q^2}{r_+^4(M,q)}-\frac{2\mu q^4}{r_+^8(M,q)}. \label{crit-cond}
\end{eqnarray}
\begin{figure*}[t!]
   \centering
  \includegraphics[width=0.5\textwidth]{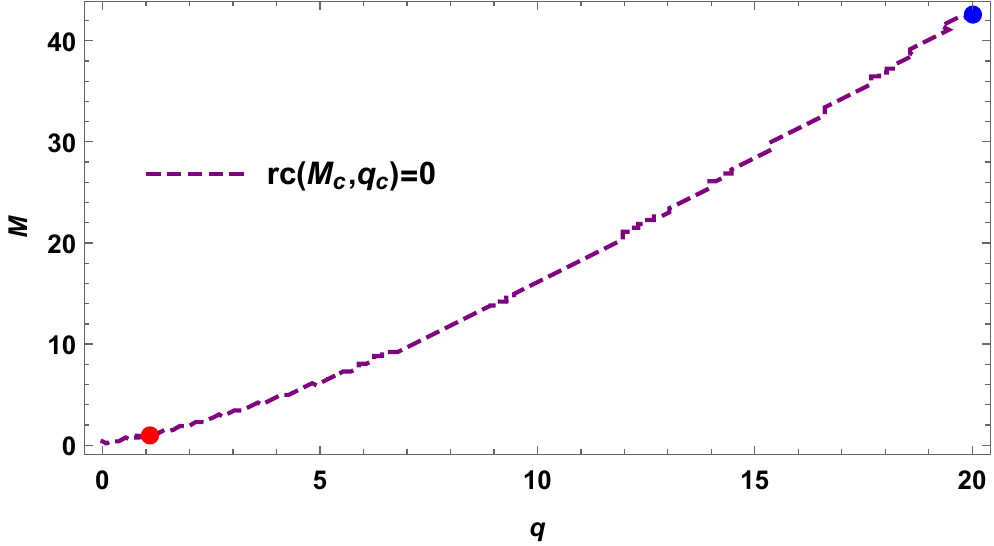}
\caption{ Resonance curve [$rc(M_c,q_c)=0$] as functions of $q_c\in[1.115,20]$ and $M_c\in[1,42.62]$ for $\alpha^-$ scalarization. It marks two critical onset points with a red dot ($q_c=1.115,M_c=1$) and a blue dot ($q_c=20,M_c=42.62)$ where the former has been  displayed in Fig. 2  as the red dot. }
\end{figure*}
The critical onset charge  and mass  are  determined by solving the resonance condition
\begin{eqnarray}
rc(M_c,q_c)=0 \to q_c=\frac{r_+^2(M_c,q_c)}{\sqrt{2\mu}}\label{res-con}
\end{eqnarray}
with  $\mu=0.3$.
The critical onset  parameter  for  $\alpha^-$ scalarization with $M_c=1$ leads to
\begin{eqnarray}
q_c=1.115.
\end{eqnarray}
Also,  $M_c=42.62$ is found for $q_c=20$.
From Eq.(\ref{res-con}), we find the resonance curve from $rc(M_c,q_c)=0$ [see Fig. 3] which describes a curve for   whole set of $\{q_c,M_c\}$.
Hence, the condition of $q> q_c$ might represent $\alpha^-$ scalarization.

We wish to describe onset scalarizations briefly.
First of all, we wish to compute  the sufficient condition for tachyonic instability given by~\cite{Dotti:2004sh}
\begin{equation}
\int_{r_+(M,q)}^\infty \Big[\frac{V_{\rm EEH}(r)}{f(r)}\Big]dr\equiv I<0,
\end{equation}
which determines $\alpha_{\rm sEEH}(M,q)$ as the sufficient instability condition.
\begin{figure*}[t!]
  \includegraphics[width=0.3\textwidth]{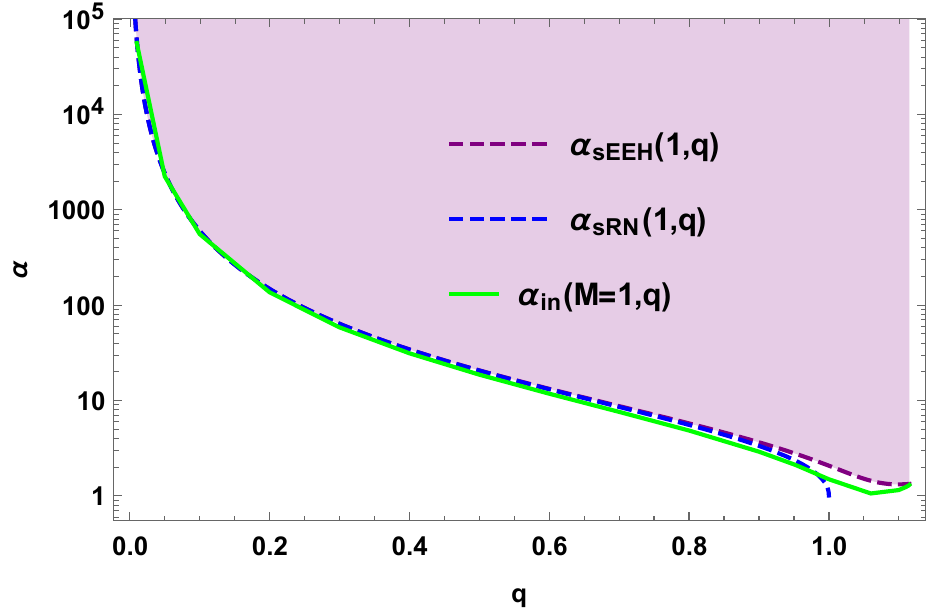}
   \hfill%
\includegraphics[width=0.3\textwidth]{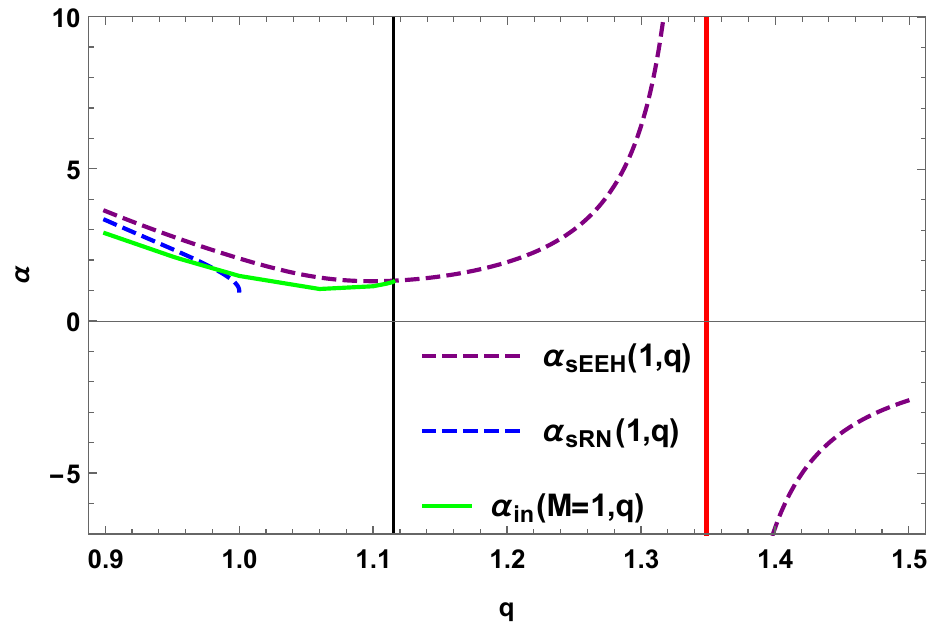}
 \hfill%
\includegraphics[width=0.3\textwidth]{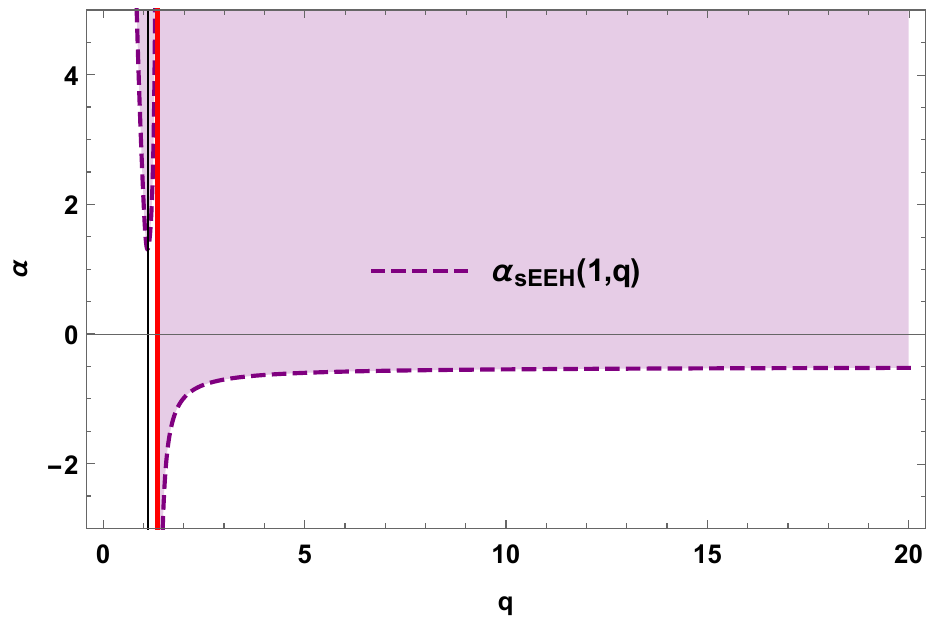}
\caption{(L) Conditions for instability $\alpha_{\rm sEEH}(1,q)$, $\alpha_{\rm sRN}(1,q)$, and $\alpha_{\rm in}(1,q)$ leading to $\alpha_{\rm sEEH}(1,q)\simeq \alpha_{\rm sRN}(1,q)\simeq \alpha_{\rm in}(1,q)$ for $q\in[0,1]$.
 The whole shaded region represents unstable region of $\alpha(1,q)\ge \alpha_{\rm sEEH}(1,q)$. One has $\alpha_{\rm sEEH}(1,q=0.5)=20.51$. (M) For $0.9\le q\le 1.5$, one has two lines such that the black hole at $q=q_c$ is an end line for $\alpha_{\rm in}(1,q)$ and the red line at $q=q_b=1.349$  denotes a blow-up line for $\alpha_{\rm sEEH}(1,q)$.
 (R) For $q> q_b$,  $\alpha_{\rm sEEH}(1,q)$ takes negative values [$\alpha_{\rm sEEH}(1,q)=-0.988(q=2), -0.523(q=20)$ ]. }\label{fig4}
\end{figure*}
We present  explicit forms $\alpha_{\rm sEEH}(M=1,q)$ and $\alpha_{\rm sRN}(M=1,q)$~\cite{Myung:2018vug} obtained from  $I=0$ as
\begin{eqnarray}
\alpha_{\rm sEEH}(1,q\in[0,\infty])&=&\frac{-1.2 q^4+7.78 q^2 r_+^4(1,q)-11.67 r_+^5(1,q)}{q^2[q^2-3.89  r_+^4(1,q)]}, \label{alp-1} \\
 \alpha_{\rm sRN}(1,q\in[0,1])&=&-2+\frac{3r_{\rm RN+}(1,q)}{q^2}, \label{alp-2}
\end{eqnarray}
which are depicted in  Fig. 4. For $0<q\le 1$, one finds that $\alpha_{\rm sEEH}(1,q) \simeq \alpha_{\rm sRN}(1,q)$. For $q> q_c=1.115$, $\alpha_{\rm sEEH}(1,q)$  blows up at $q=q_b=1.349$ and then, it takes  negative values.
This explains pictorially why  $\alpha^+$ and $\alpha^-$ scalarizations should be  present when considering  $q$-dependent scalarization of EEHBHs. There is a gap between $q=q_c$ and $q=q_b$ because we use $\alpha_{\rm sEEH}(1,q)$ [see Fig. 4(M)].

To obtain the instability condition $\alpha_{\rm in}(M,q)$, one may use  the spatially regular  scalar configurations (scalar clouds) which  can be obtained  from   Eq.(\ref{mode-d}) without time-dependence  by adopting  the WKB method~\cite{Hod:2019ulh}.
A standard WKB method  could be used to obtain the bound states of $\varphi_{00}$, yielding  the quantization condition
\begin{equation}
\int^{r_*^{\rm out}}_{r_*^{\rm in}}dr_* \sqrt{-V_{\rm EEH}(r_*)}=\Big(n-\frac{1}{4}\Big)\pi,\quad n=1,2,3,\cdots. \label{wkb1}
\end{equation}
Here, $r_*^{\rm out}$ and $r_*^{\rm in}$ are two turning points satisfying $V_{\rm EEH}(r_*^{\rm out})=V_{\rm EEH}(r_*^{\rm in})=0$.
We may  rewrite   Eq.(\ref{wkb1})  as
\begin{equation}
\int^{r_{\rm out}}_{r_{\rm in}}dr \frac{\sqrt{-V_{\rm EEH}(r)}}{f(r)}=\Big(n-\frac{1}{4}\Big)\pi. \label{wkb2}
\end{equation}
Radial turning points $(r_{\rm out}$ and $r_{\rm in})$ are determined by imposing the two conditions
\begin{equation}
f(r_{\rm in})=0,\quad \frac{2M}{r^3_{\rm out}}-\frac{(\alpha+2)q^2}{r^4_{\rm out}}+\frac{2\mu q^4(\alpha+6)}{5r^8_{\rm out}}=0,
\end{equation}
implying 
\begin{equation}
r_{\rm in}=r_+(M,q),~\quad r_{\rm out}(M,q,\alpha).
\end{equation}
For large $\alpha(r_{\rm out})$, the WKB integral (\ref{wkb2}) is  approximated by considering the mass term $m_{\rm eff}^2$ in (\ref{EEH-P}) as
\begin{equation} \label{int-e}
\sqrt{\alpha}\cdot  q\int^{\infty}_{r_+} dr \frac{\sqrt{1-\frac{2\mu q^2}{r^4}}}{r^2\sqrt{f(r)}}\equiv\sqrt{\alpha} I_n(M,q) =\Big(n+\frac{3}{4}\Big)\pi,\quad  n=0,1,2,\cdots,
\end{equation}
which could be integrated numerically to yield positive quantities 
\begin{equation} \label{alphan}
\alpha_{\rm in,n}(M,q)=\Big[\frac{\pi(n+3/4)}{I_n(M,q)}\Big]^2.
\end{equation}
We plot $\alpha_{\rm in}(M=1,q)[\equiv\alpha_{\rm in,n=0}(1,q)]$ in Fig. \ref{fig4}, which is available  for $0<q< q_c=1.115$.
It is noted  that $\alpha_{\rm in}(1,q)[ \simeq \alpha_{\rm sEEH}(1,q)]$ is a decreasing function of $q$ and  others of $\alpha_{\rm in,n\not=0}(1,q) $ may be  used to estimate  branch points:  $\alpha_{\rm in,0}(1,0.5)=18.51$,  $\alpha_{\rm in,1}(1,0.5)=100.77 $, and  $\alpha_{\rm in,2}(1,0.5)=248.84$. However, for $q>q_c$, one cannot determine  $\alpha_{\rm in,n}(M=1,q)$ because the integration in Eq.(\ref{int-e}) is not properly defined.
\begin{figure*}[t!]
   \centering
  \includegraphics[width=0.4\textwidth]{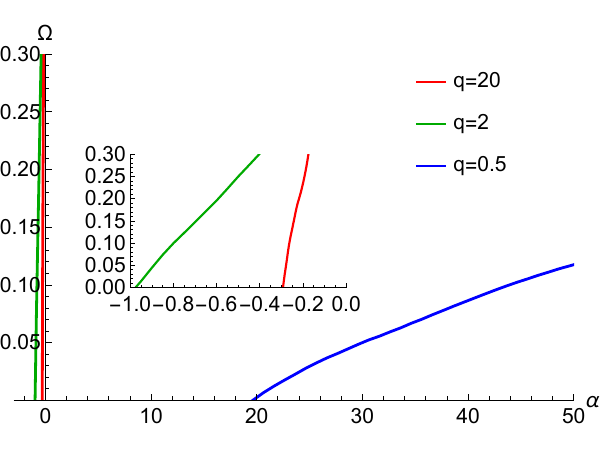}
  \hfill%
\includegraphics[width=0.4\textwidth]{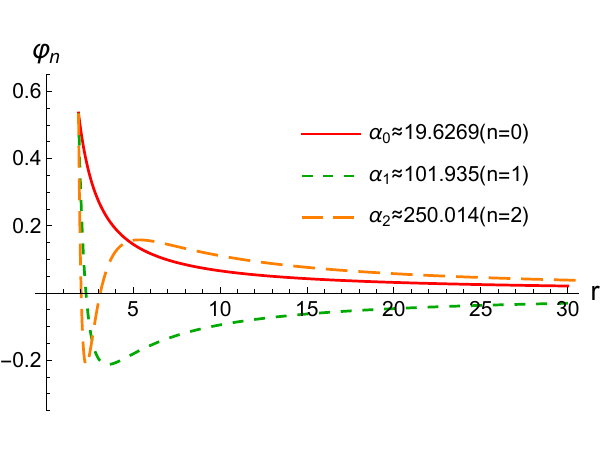}
\caption{
(Left) Three curves of $\Omega$ in $e^{\Omega t}$ as a function of $\alpha$ are used to determine the thresholds of tachyonic instability $[\alpha_{\rm th}(1,q)]$ around the EEHBHs.
We find that  $\alpha_{\rm th}(1,q) = 19.63(q=0.5),~ -0.9762(q=2),~ -0.2927(q=20)$ when three curves cross the $\alpha$-axis.
 (Right) Radial profiles of the static scalar $\varphi(r)=u(r)/r$ as function of $r\in[r_+=1.87,30]$ denote the first three  scalar clouds for $q=0.5<q_c$.
These solutions  $\varphi_n(r)$ are classified by the node number ($n=0,1,2$) and will be seeds for generating $n=0,1,2$ branches of scalarized EEHBHs.  }
\end{figure*}

Importantly, to determine the threshold of tachyonic instability $\alpha_{\rm th}(M=1,q)$ precisely, we have to solve Eq.(\ref{mode-d}),
which  allows an exponentially growing mode of  $e^{\Omega t}(\omega_i=\Omega>0) $ as  an unstable mode for $\omega=\omega_r+i\omega_i$ with $\omega_r=0$.
Two boundary conditions are required:  normalizable
solution of $u(\infty)\sim e^{-\Omega r_*}$  at infinity  and
 power solution of $u(r_+)\sim \left(r-r_+\right)^{\Omega r_+}$  in the near-horizon.
We find from (Left) Fig. 5 that  the threshold ($\Omega=0$) of tachyonic  instability  can be read off as  $\alpha_{\rm th}(1,q) = 19.63(q=0.5),~ -0.98(q=2),~ -0.293(q=20)$ which are less than those of $\alpha_{\rm sEEH}(1,q)$.
This shows clearly that  the EEHBH is unstable for $\alpha(1,q)>\alpha_{\rm th}(1,q)$, while it is stable for  $\alpha(1,q)<\alpha_{\rm th}(1,q)$. 
\begin{figure*}[t!]
   \centering
  \includegraphics[width=0.28\textwidth]{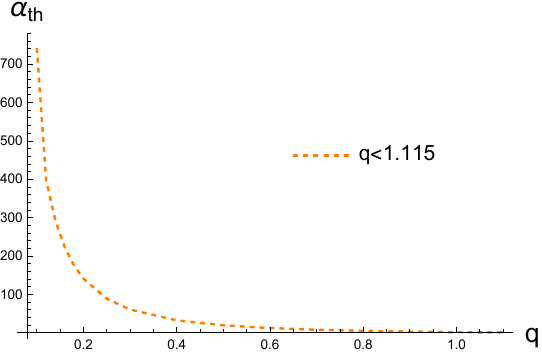}
  \hfill%
\includegraphics[width=0.28\textwidth]{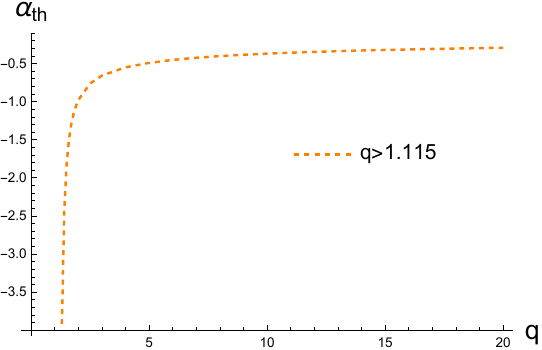}
\hfill%
\includegraphics[width=0.28\textwidth]{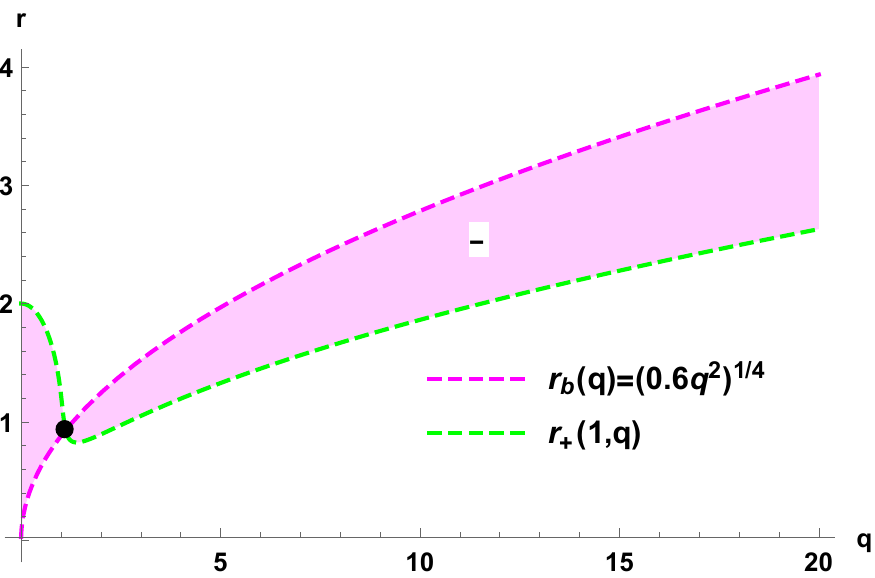}
\caption{ Two threshold curves $\alpha_{th}(q)$ as functions of $q$ and negative region for obtaining $\alpha_{\rm in,n}(1,q>q_c)$. 
(L)  $\alpha_{th}(q)>0~(0<q<q_c=1.115)$ to have $\alpha^+$ scalarization.
(M) $\alpha_{th}(q)<0~(q_c<q\le20)$ to get $\alpha^-$ scalarization. (R) A dangerous negative region for computing $\alpha_{\rm in,n}(1,q)$ with $q=2,20$. This corresponds to $1-\frac{2\mu q^2}{r^4}<0$ in Eq.(\ref{int-e}). The black dot denotes the critical onset point ($q_c,r_+(1,q_c))$. }
\end{figure*}
 As is shown in  (L,M) Fig. 6, two threshold curves $\alpha_{th}(q)$ are quite different:  one has  $\alpha_{th}(q)>0$ for $0<q<q_c=1.115$  to have $\alpha^+$ scalarization, 
whereas   $\alpha_{th}(q)<0$ is shown  for $q_c<q\le20$  to get $\alpha^-$ scalarization.
This clearly shows a relation between critical onset charge $q_c$ and threshold curves $\alpha_{th}(q)$. 

For $\alpha^+$ scalarization with $0<q< q_c$,  the other way to determine   $\alpha_{\rm th}(1,q)$  is to solve the  static linearized  equation directly.
Also, this can be used to construct the scheme for infinite branches ($n=0,~ 1,~ 2,\cdots$)  of scalarized EEHBHs.
For this purpose, the static linearized equation (\ref{per-eq}) for $s$-mode $\delta \varphi=\varphi(r)\cdots$  is introduced as 
\begin{equation} \label{ssclar-eq}
\frac{1}{r^2}\frac{d}{dr}\Big[r^2f(r)\frac{d\varphi(r)}{dr}\Big]-m^2_{\rm eff} \varphi(r)=0,
\end{equation}
which defines an eigenvalue problem: requiring an asymptotically vanishing and smooth scalar field
selects a discrete set of $n=0,1,2,\cdots$ and simultaneously
determines the correct bifurcation points $\{\alpha_n(1,q)\}$.
We confirm that  $\alpha_{\rm th}(M=1,q=0.5)=\alpha_{0}(1,0.5)$.
We plot $\varphi_n(r)$ with $q=0.5<q_c$ as a function of $r$ for the lowest  three cases of  $n=0(\alpha_0=19.6269),~n=1(\alpha_1=101.935),~n=2(\alpha_2=250.014)$ whose scalar clouds are depicted in (Right) Fig. 5.
It is worth noting    that the  scalar cloud $\varphi_0(r)$ without zero crossing will develop the $n=0$  branch of scalarized EEHBHs existing for  $\alpha\ge \alpha_0$. On the other hand, 
 two scalar clouds  $\varphi_1(r)$ and $\varphi_2(r)$ with zero crossings will develop the $n=1$ and $n=2$  branches of scalarized EEHBHs  existing for  $\alpha\ge \alpha_1$ and $\alpha\ge\alpha_2$, respectively.

 Finally, for $\alpha^-$ scalarization with $q>q_c$, we solve the
static equation (26) directly for $q=2$ and $q=20$ with negative
eigenvalues. However, within the parameter ranges explored, we do not
find additional scalar-cloud solutions with nodes. This result is
consistent with the absence of $\alpha_{{\rm in},n}(1,q>q_c)$ caused
by the negative region outside the event horizon shown in Fig.~6(R).
Accordingly, in our numerical construction below, only one scalarized
branch is obtained for each of $q=2$ and $q=20$ within the explored
parameter ranges.

\section{Construction of scalarized EEHBHs}

 Infinite  branches of scalarized EEHBHs were generated from the onset of $\alpha^+$ scalarization $\{\varphi_n(r)\}$ in the unstable region of EEHBHs [$\alpha(1,q)\ge \alpha_{\rm th}(1,q)$] for $0<q< q_c$.
For $\alpha^-$ scalarization with $q>q_c$, scalarized EEHBH solutions can be constructed directly.

In this section, we wish to construct  all scalarized EEHBHs  by solving full equations.
For this purpose,  we introduce  the metric and fields as~\cite{Herdeiro:2018wub}
\begin{eqnarray}\label{nansatz}
ds^2_{\rm SEEH}&=&-N(r)e^{-2\delta(r)}dt^2+\frac{dr^2}{N(r)}+r^2(d\theta^2+\sin^2\theta d\hat{\varphi}^2) \nonumber \\
N(r)&=&1-\frac{2m(r)}{r},\quad \phi=\phi(r),\quad A= A_{\hat{\varphi}} d\hat{\varphi}.
\end{eqnarray}
Substituting the gauge field ansatz into Eq.(\ref{M-eq}), one finds  a magnetic  potential $A_{\hat{\varphi}}=-q \cos \theta$, leading to $F_{\theta \hat{\varphi}}=q\sin\theta$ and $\mathcal{F}=2q^2/r^4$.
This means that it is unnecessary to consider an approximate solution for $A_{\hat{\varphi}}$. Thus, this completes the solution of the Maxwell equation when considering the magnetic charge.
Plugging (\ref{nansatz}) into Eqs.(\ref{equa1}) and (\ref{s-equa}),  three equations for $m(r),~\delta(r),$ and $\phi(r)$ are given by
\begin{eqnarray}
&&e^{-\alpha\phi^2(r)}\Big(q^2-\frac{2\mu q^4}{r^4}\Big)-2r^2m'(r)+r^3\Big(r-2m(r)\Big)\phi'^2(r)=0, \label{neom1}\\
&&\delta'(r)+r\phi'^2(r)=0, \label{neom2}\\
&&\frac{\alpha \Big(q^2-\frac{2\mu q^4}{r^4}\Big)\phi(r)e^{-\alpha\phi^2(r)}}{r^2}-2\Big[m(r)+rm'(r)-r\Big]\phi'(r) \nonumber \\
&&\quad\quad-r\Big(r-2m(r)\Big)\Big[\delta'(r)\phi'(r)-\phi''(r)\Big]=0. \label{neom3}
\end{eqnarray}
Here, the prime ($'$) denotes differentiation with respect to $r$. It is  noted that Eq.(\ref{neom1}) reduces to Eq.(\ref{mass-eq}) for $\phi(r)=0$.
\begin{figure*}[t!]
   \centering
  \includegraphics[width=0.3\textwidth]{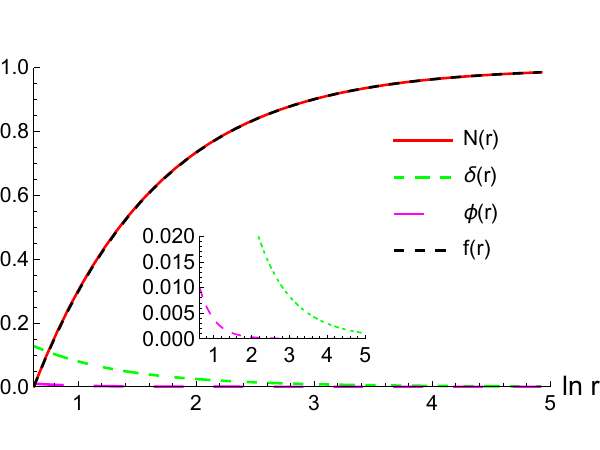}
  \hfill%
\includegraphics[width=0.3\textwidth]{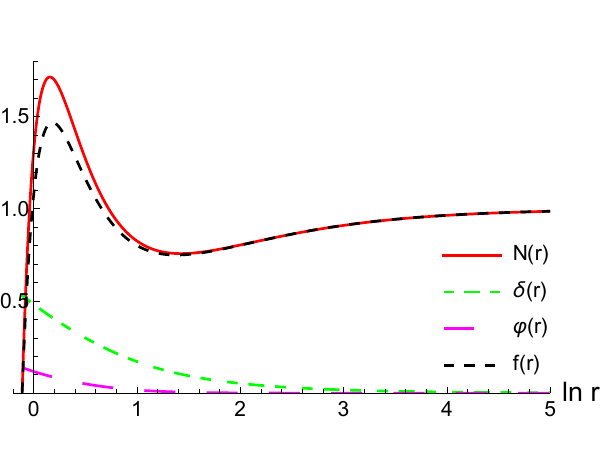}
 \hfill%
\includegraphics[width=0.3\textwidth]{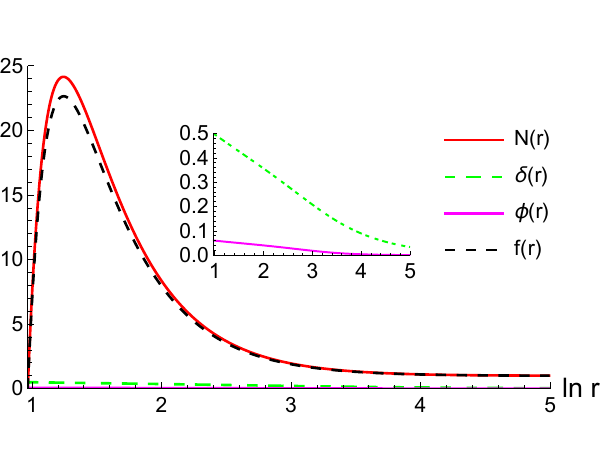}
\caption{Graph  of scalarized EEHBH  solutions. It shows metric functions $N(r)$, $\delta(r)$, and $f(r)$ for  EEHBH, and scalar hair $\phi(r)$. (Left) $q = 0.5$, $M=1$, and $\ln{r_{+}}=0.625$ for $\alpha = 25$ in the $n = 0$ branch of $\alpha \geq 19.63$.
(Middle) $q=2$, $M=1$, and $\ln r_+=-0.1115$ for $\alpha=-0.90$ on the branch obtained for $\alpha\geq-0.9762$.
(Right)  $q=20$, $M=1$, and $\ln r_+=0.9678$ for $\alpha=-0.290$ on the branch obtained for $\alpha\geq-0.2927$.}
\end{figure*}
Considering  the existence of a single horizon located at $r=r_+$,   an
approximate solution to Eqs.(\ref{neom1})-(\ref{neom3})  takes the following form near the horizon:
\begin{eqnarray}
m(r)&=&\frac{r_+}{2}+m_1(r-r_+)+\cdots,\quad
\delta(r)=\delta_0+\delta_1(r-r_+)+\cdots,\label{aps-1}\\
\phi(r)&=&\phi_0+\phi_1(r-r_+)+\cdots,\label{aps-2}
\end{eqnarray}
where  three coefficients are given by
\begin{eqnarray}\label{ncoef}
&&m_1=\frac{e^{-\alpha\phi_0^2}}{2r_+^2}\Big(q^2-\frac{2\mu q^4}{r_+^4}\Big),\quad \delta_1=-r_+\phi_1^2,\quad \phi_1=\frac{\alpha \phi_0e^{-\alpha\phi_0^2}\Big(q^2-\frac{2\mu q^4}{r_+^4}\Big) }{r_+(2m_1-1)}.
\end{eqnarray}
Importantly, we observe that these all disappear if $2\mu q^2=r_+^4(q=q_c)$.  
Here, two parameters of $\phi_0=\phi(r_+,\alpha)$ and $\delta_0=\delta(r_+,\alpha)$ will be
determined when matching with an asymptotically flat solution in the far-region
\begin{eqnarray}\label{ncoef}
&&m(r)=M-\frac{q^2+q_s^2}{2r}+\cdots,\quad
\delta(r)=\frac{q_s^2}{2r^2}+\cdots,\quad
\phi(r)=\phi_\infty+\frac{q_s}{r}+\cdots.
\end{eqnarray}
 Here,  $q_s$ will be  a secondary  scalar charge, in addition to the ADM mass $M$, and the magnetic charge $q$.

Regarding the explicit scalarized EEHBH solutions with $q=0.5$ ($n=0$ branch), $2$, and $20$, we present the
numerical solutions in Fig 7. 
Further, we need  to explore  hundreds of numerical solutions depending on $\alpha$ for each $q$ to perform their stability analysis of scalarized EEHBHs.
 To clarify the global structure of the scalarized solutions, we further
display the scalarized black-hole branches in the $(\alpha,\phi_0)$
plane with the horizon  scalar
\begin{equation}
 \phi_0=\phi(r_+).
\end{equation}
\begin{figure}[ht]
\centering
\includegraphics[width=0.3\textwidth]{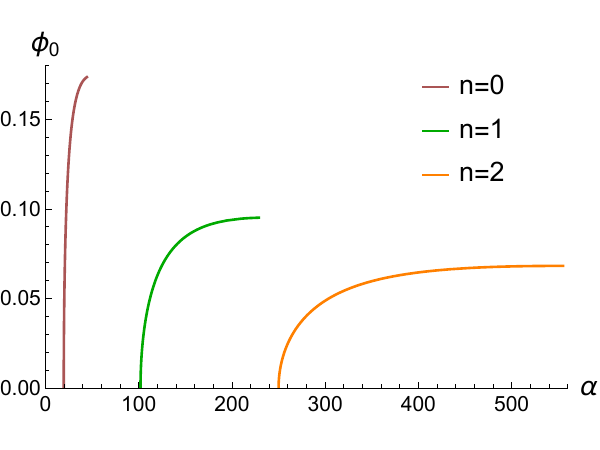}
 \hfill%
\includegraphics[width=0.3\textwidth]{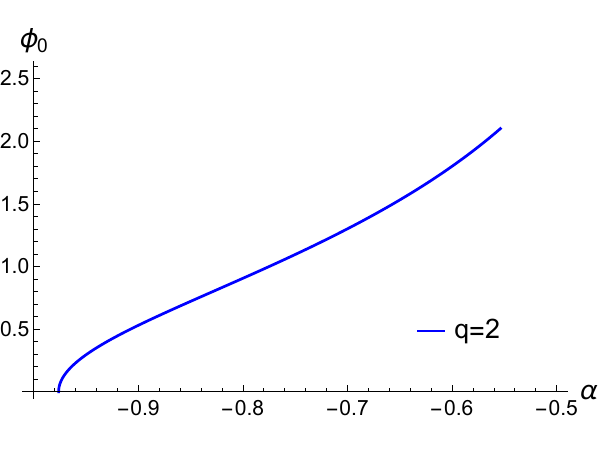}
 \hfill%
\includegraphics[width=0.3\textwidth]{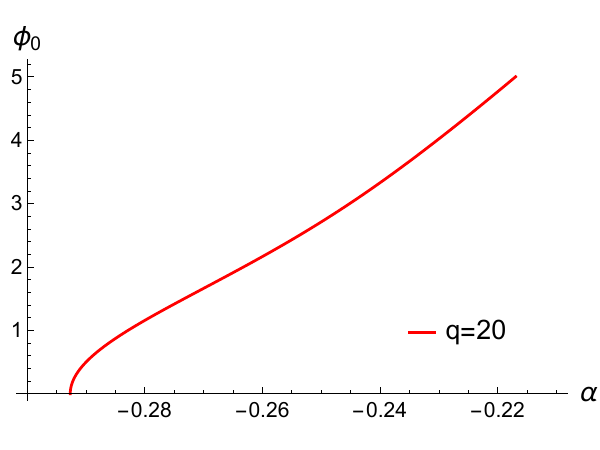}
\caption{
Scalarized EEHBH branches in the $(\alpha,\phi_0)$ plane.
The left panel shows the $n=0,1,2$ branches for $q=0.5$, while
the middle and right panels show the branches obtained for $q=2$
and $q=20$, respectively, within the explored parameter ranges.
Each branch bifurcates from the unstable EEHBH at the threshold
coupling $\alpha_{\rm th}(q)$ when $\phi_0=0$.}

\label{fig:branches}
\end{figure}
As shown in Fig.~\ref{fig:branches}, the scalarized branches emerge
continuously from the unstable EEHBHs at
$\alpha=\alpha_{\rm th}(q)$. For $q=0.5$, the first three branches are labeled
by node number  $n=0,1,2$. In contrast, for $q=2$ and $q=20$, only one scalarized branch is
obtained for each charge within the parameter ranges explored.
The bifurcation points and the largest values of $\alpha$ explored
along these branches are summarized in Table~\ref{tab:branch-range}.
\begin{table}[ht]
\centering
\caption{Bifurcation points and the largest values of $\alpha$ explored for the scalarized EEHBH branches.}
\label{tab:branch-range}
\begin{tabular}{c|c|cc}
\hline
$q$ & Branch & Bifurcation point & Largest value explored \\
\hline
0.5 & $n=0$ & $19.63$ & $44.44$ \\
0.5 & $n=1$ & $101.9$ & $228.8$ \\
0.5 & $n=2$ & $250.0$ & $555.1$ \\
2   & $n=0$ & $-0.9762$ & $-0.5538$ \\
20  & $n=0$ & $-0.2927$ & $-0.2170$ \\
\hline
\end{tabular}
\end{table}
Near the bifurcation points, $\phi_0$ approaches zero continuously
and the scalarized solutions reduce to the corresponding 
EEHBHs. The largest values of $\alpha$ explored in our
numerical computation are listed in Table~\ref{tab:branch-range}
but they should not necessarily be interpreted as physical endpoints of
the branches.  \\ Finally,  in deriving physical quantities of this section, numerical method and convergence test are displayed in Appendix A.

\section{Stability of scalarized EEHBHs for $q=0.5$ ($n=0$),
2, and 20}

First of all, we would like to mention   that the stability analysis for scalarized EEHBHs is an important issue
since it determines their viability in representing realistic astrophysical configurations.
The conclusions on  the stability of the scalarized EEHBHs  under  radial perturbations might  be reached by examining
the qualitative behavior of their potentials  and  by finding  exponentially growing (unstable) modes for $s$-mode scalar  perturbation.

For this purpose,  we wish  to  introduce the radial perturbations around scalarized EEHBHs (belonging to polar-perturbations)  as
\begin{eqnarray}
&&ds_{\rm rp}^2=-N(r)e^{-2\delta(r)}(1+\epsilon H_0)dt^2+\frac{1}{N(r)}(1+\epsilon H_2)dr^2
+r^2(d\theta^2+\sin^2\theta d\hat{\varphi}^2),\nonumber\\
&&\phi(t,r)=\phi(r)+\epsilon\delta\tilde{\phi}(t,r), \label{p-metric}
\end{eqnarray}
where $N(r)$, $\delta(r)$, and  $\phi(r)$ represent the scalarized
EEHBH background, whereas
$H_0(t,r)$, $H_2(t,r)$, and $\delta\tilde{\phi}(t,r)$
denote three perturbations  around the scalarized EEHBH background. 
From now on, we consider  the $l=0$($s$-mode)
scalar mode by neglecting   higher angular momentum modes $(l\neq0)$. In this
case, the other two perturbations ($H_0,H_2$)  become redundant perturbations.
Considering  a decoupling process, one finds a linearized scalar equation.
 For a detailed derivation, see Appendix B.

Its time-dependent  linearized equation is given by 
\begin{equation} \label{tmode-d}
\frac{\partial^2 \delta\tilde{\varphi}(t,r_*)}{\partial r_*^2} -\frac{\partial^2\delta\tilde{\varphi}(t,r_*)}{\partial t^2}=V_{\rm sEEH}(r)\delta\tilde{\varphi}(t,r_*)
\end{equation}
where $r_*$ is the tortoise coordinate defined by
\begin{eqnarray}
\frac{dr_*}{dr}=\frac{e^{\delta(r)}}{N(r)}.
\end{eqnarray}
Here, its potential is given by
\begin{align} \label{sc-poten}
V_{\rm sEEH}(r,q,\alpha) &= \frac{N}{r^4} e^{-2 \delta - \alpha \phi^2} 
\Bigg[ q^2 \bigl(1 - \frac{2 q^2 \mu}{r^4} \bigr) 
\Bigl( 2 \alpha^2 \phi^2 
- 4 r \alpha \phi \phi' 
+ r^2 \phi'^2  -1 - \alpha \Bigr) \nonumber \\
& \quad - e^{\alpha \phi^2} r^2 \bigl(  2 r^2 \phi'^2  + N -1\bigr) \Bigg]
\end{align}
Considering the separation of variables
\begin{eqnarray} \label{sps}
\delta\tilde{\phi}(t,r)=\frac{\tilde{\varphi}(r)e^{-i\omega t}}{r},
\end{eqnarray}
we obtain the time-independent Schr\"odinger-type equation for an $s$-mode scalar perturbation
\begin{eqnarray} \label{tiS-eq}
\frac{d^2\tilde{\varphi}(r)}{dr_*^2}+\Big[\omega^2-V_{\rm sEEH}(r,q,\alpha)\Big]\tilde{\varphi}(r)=0.
\end{eqnarray}
{\bfseries
The complete spectral problem is handled by solving Eq.~(41) together
with the quasinormal mode (QNM) boundary conditions. Considering the time
dependence $e^{-i\omega t}$, they are  a purely ingoing wave at
the event horizon and a purely outgoing wave at spatial infinity,
\[
\tilde{\phi}\sim e^{-i\omega r_*},
\qquad r_*\rightarrow-\infty,
\]
and
\[
\tilde{\phi}\sim e^{+i\omega r_*},
\qquad r_*\rightarrow+\infty,
\]
respectively.
\begin{figure*}[t!]
\centering
\includegraphics[width=0.3\textwidth]{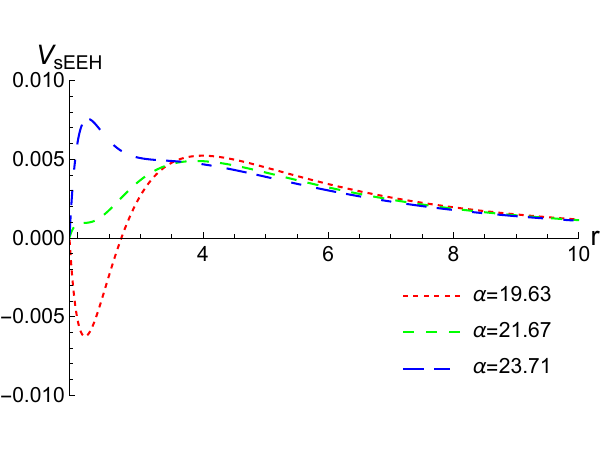}
 \hfill%
\includegraphics[width=0.3\textwidth]{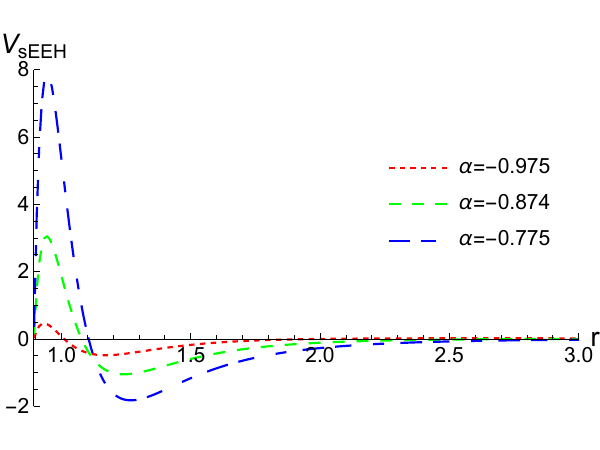}
 \hfill%
\includegraphics[width=0.3\textwidth]{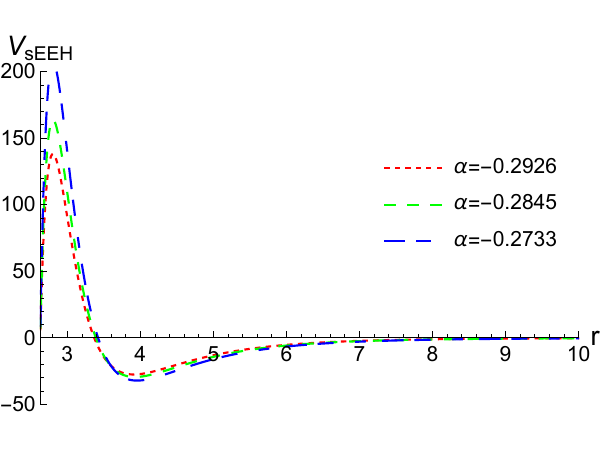}
\caption{ (Left) Three scalar potentials $V_{\rm sEEH}(r,q=0.5,\alpha)$  for $\alpha=19.63,21.67,23.71$ around the $n = 0$ branch.  (Middle)
Three scalar potentials $V_{\rm sEEH}(r,q=2,\alpha)$  for $\alpha=-0.975,-0.874,-0.775$. (Right) Three scalar potentials $V_{\rm sEEH}(r,q=20,\alpha)$  for $\alpha=-0.2926,-0.2845,-0.2733$. 
Even though they  contain small negative regions in the near horizon, these  turn out to be  stable black holes.}
\label{9}
\end{figure*}
To obtain the stability criterion, we choose the purely imaginary frequency,
$\omega=i\Omega$. In this case, $\Omega<0$ corresponds
to an exponentially decaying mode $e^{\Omega t}$,
whereas $\Omega>0$ describes an exponentially growing mode.
\par}
\begin{figure*}[t!]
   \centering
  \includegraphics[width=0.3\textwidth]{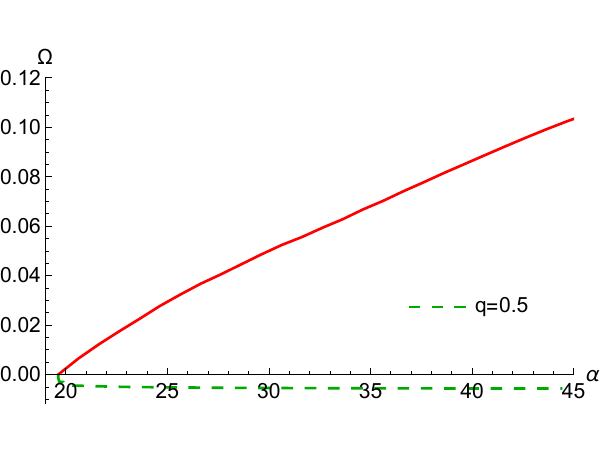}
   \hfill%
\includegraphics[width=0.3\textwidth]{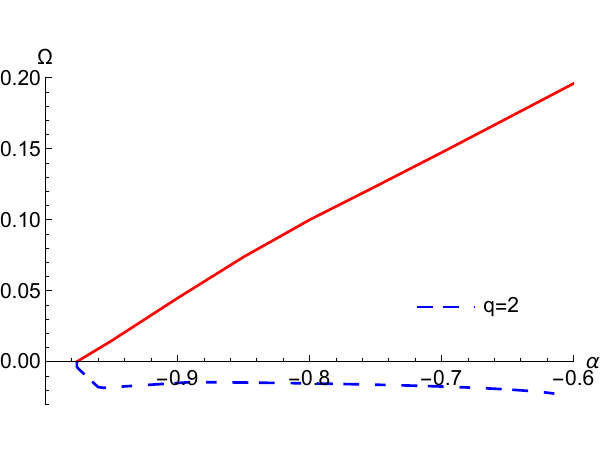}
 \hfill%
\includegraphics[width=0.3\textwidth]{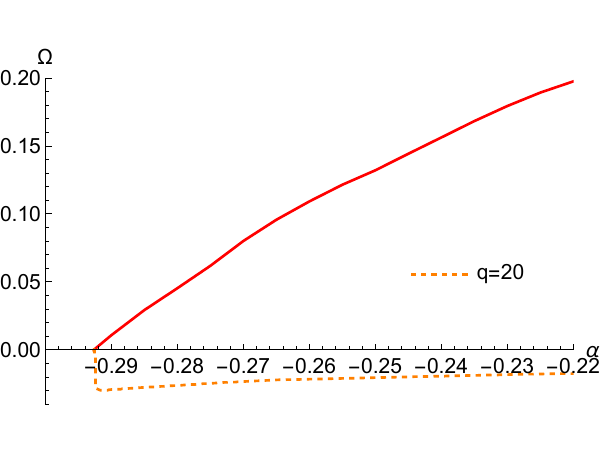}
\caption{ \textbf{Purely imaginary QNM frequency $\omega=i\Omega$
as a function of $\alpha$ for the $l=0$ scalar mode with
$q=0.5$ ($n=0$), $2$, and $20$. Negative $\Omega$ corresponds
to exponentially damped modes.} The dotted curves start from
$\alpha_{n=0}=19.63$, $-0.9762$, and $-0.2927$, respectively.
The red lines represent unstable EEHBHs [see (Left) Fig.~5].}
\label{10}
\end{figure*}
We check first that $V_{\rm sEEH}(r,q,\alpha)$ with
$\delta(r)=\phi(r)=0$ and $N(r)\to f(r)$ reduces to
$V_{\rm EES}(r,M,q,\alpha)$ in Eq.~(\ref{EEH-P}).
At this stage, we observe the potential $V_{\rm sEEH}(r,q,\alpha)$.
We display three scalar potentials $V_{\rm sEEH}(r,q,\alpha)$
for $q=0.5(n=0),2,20$ in Fig.~9, showing small negative regions
for $q=0.5$ near the horizon.
However, this does not imply that the $n=0$ branch is unstable
against the $s$-mode scalar perturbation because the sufficient
condition for instability~\cite{Dotti:2004sh} is given by
$\int_{r_+}^\infty dr[e^\delta V_{\rm sEEH}(r,q=0.5,\alpha)/N]<0$.
It suggests that the $n=0$ branch with $q=0.5$ may be stable
against the $s$-mode scalar perturbation.
Actually, we have to solve Eq.~(\ref{tmode-d}) using
Eq.~(\ref{sps}) with $\omega=i\Omega$ ($e^{\Omega t}$)
to analyze its stability.
\textbf{As shown in Figs.~\ref{10} and \ref{11}, the purely imaginary
frequencies satisfy $\Omega<0$ for $q=0.5$ ($n=0$ branch),
$q=2$, and $q=20$. These negative values correspond to
exponentially decaying modes. Here, the three red curves in
Fig.~\ref{10}, starting at $\alpha_{\rm th}=19.63$ ($q=0.5$),
$-0.9762$ ($q=2$), and $-0.2927$ ($q=20$), correspond to
$\Omega>0$ and hence to exponentially growing modes of the
unstable EEHBHs. Furthermore, Fig.~\ref{11} provides a zoomed-in
view for the $q=2$ and $q=20$ cases. Our radial perturbation
analysis indicates that the scalarized branches are stable against
the $l=0$ modes considered here. Representative convergence tests
for these frequencies are presented in Appendix A.}

{\bfseries In addition,  we expect the excited branches ($n\geq1$) for $q=0.5$ to be
unstable, similarly to the model with an exponential scalar
coupling to the Maxwell term only~\cite{Zhang:2025msi}.\par}

\begin{figure*}[t!]
   \centering
  \includegraphics[width=0.4\textwidth]{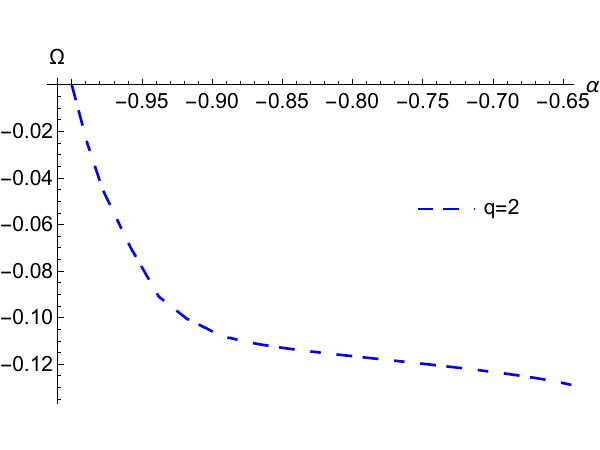}
  \hfill%
\includegraphics[width=0.4\textwidth]{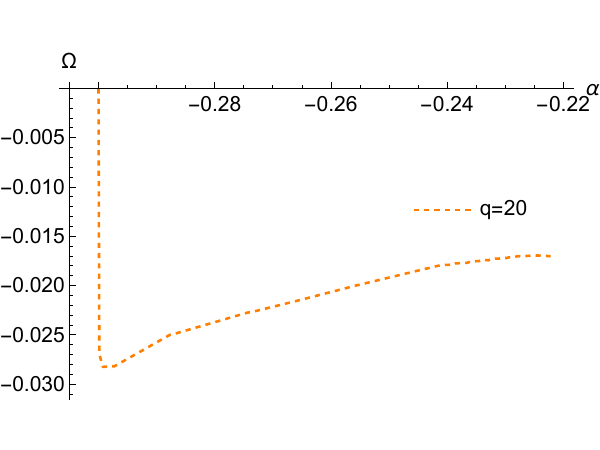}
\caption{\textbf{Zoomed-in view of the purely imaginary QNM
frequencies $\omega=i\Omega$ for 
$q=2$ and $q=20$. The negative values of $\Omega$ describe exponentially decaying modes.} }
\label{11}
\end{figure*}

\section{Discussions }
Scalarization provided a dynamical mechanism for the formation of hairy black holes and  was of great significance for understanding the interaction between gravity and matter. It was mainly determined by potentials for the minimal coupling to gravity and  forms of the coupling function to  matter for nonminimal couplings. 

Yajima and Tamaki have obtained  the Einstein-Euler-Heisenberg (EEH) black hole solution described by mass ($M$), charge ($q$), and the EH parameter ($\mu$) when considering Einstein-Euler-Heisenberg theory including  the NED term ($\mathcal{F}^2$)~\cite{Yajima:2000kw}.  One feature of this solution is that it can possess the single horizon without limitation on the charge $q>0$ for $\mu> 0.08$. The other case of $\mu\le 0.08$ has multiple horizons.

A negative potential-induced scalarization of the EEH black hole with single horizon ($\mu=0.3$) was investigated  with unlimited charge $q$ in the EEH-minimally coupled scalar theory~\cite{Guo:2025ksj}. It turned out that the single branch of scalarized black holes was unstable against  radial perturbations, but with  $M=1/2$, the scalar charge $q_s$ exhibits a primary scalar hair for $ q < 1/2$, whereas  it becomes a constant (secondary scalar hair) for $q>1/2$.
Also,  spontaneous scalarizations of the  EEH  black hole with single horizon  were  performed for $q=0.5,~2,~20$ in the EEHS theory by introducing an exponential coupling  to the Maxwell term only~\cite{Zhang:2025msi}.  In this case, there was some difference in the onset of scalarization between $q\le1$ and $q>1$ but  all $n=0$ branches are stable against the radial perturbations. 

In the present work, we have carried out charge-dependent scalarization of the  EEH black hole with single horizon  in the EEHS theory by introducing an exponential scalar coupling to the Maxwell and NED terms.
Spontaneous scalarization ($\alpha^+$) of this black hole was performed   for  charge $0<q< q_c=1.115$ and positive $\alpha$, whereas  its new scalarization  ($\alpha^-$) occurred for $q> q_c$ and negative $\alpha$.  
The former case of $q=0.5$  implies infinite branches of scalarized EEHBHs, but its fundamental branch ($n=0$) is stable against the $l=0$ modes considered here. On the other hand, for $q=2$ and $q=20$, one stable scalarized
branch was found for each charge within the explored parameter ranges.

\textbf{Consequently, this work has shown clearly two distinct scalarizations
depending on the charge ($q$) when including a single exponential
scalar coupling to the combined Maxwell and NED terms.} This result contrasts with the spontaneous scalarization  obtained from the Einstein-Maxwell-scalar theory with the same exponential scalar coupling to the Maxwell term only~\cite{Herdeiro:2018wub,Myung:2018vug}, where over-charge ($q>1$) appeared in spontaneous scalarization.

 \vspace{1cm}

{\bf Acknowledgments}

 Y.S.Myung was supported by the National Research Foundation of Korea (NRF) grant funded by the Korea government(MSIT) (RS-2022-NR069013). D.C. Zou was  supported by the National Natural Science Foundation of China (NNSFC) (Grant No.12365009).
 
 \vspace{1cm}
\newpage 

\appendix
\section{Numerical method and convergence tests}

We  construct  scalarized EEHBHs by  solving
Eqs.~(28)--(30) numerically together with a shooting method. Near the event horizon, the
metric functions and scalar field are expanded as
\begin{equation}
m(r)=\sum_{i=0}^{N_h}m_i(r-r_+)^i,\qquad
\delta(r)=\sum_{i=0}^{N_h}\delta_i(r-r_+)^i,\qquad
\phi(r)=\sum_{i=0}^{N_h}\phi_i(r-r_+)^i.
\end{equation}
where $N_h=3$ is adopted for  our numerical calculations. To avoid the
coordinate singularity at the horizon, the numerical integration starts at
\begin{equation}
 r_a=(1+10^{-3})r_+
\end{equation}
and extends to
\begin{equation}
 r_{\rm max}=e^6\simeq 403.43 .
\end{equation}
At large $r$, the numerical solutions are matched with  the asymptotic expansions
\begin{equation}
 m(r)=M-\frac{q^2+q_s^2}{2r}+\cdots,\qquad
 \delta(r)=\frac{q_s^2}{2r^2}+\cdots ,\qquad 
 \phi(r)=\phi_\infty+\frac{q_s}{r}+\cdots .
\end{equation}
The shooting parameters are determined by imposing
\begin{equation}
 \phi_\infty=0,\qquad \delta_\infty=0 ,
\end{equation}
while the scalar charge $q_s$ can be  extracted from the coefficient of the
$1/r$ term in the asymptotic scalar. Here,  we note that  an adaptive integration scheme
is employed  with target numerical tolerances of $10^{-8}$ for the radial
integration and $10^{-5}$ for the shooting iterations.

\textbf{To find the radial perturbation spectrum, we solve Eq.~(41) by setting
$\omega=i\Omega$ and imposing the QNM boundary
conditions: a purely ingoing wave at the event horizon and
a purely outgoing wave at spatial infinity. We note that the threshold coupling
$\alpha_{\rm th}(q)$ was  separately determined by the marginal
static mode with $\Omega=0$.}

To examine the dependence on the finite numerical domain, we vary the
outer boundary as
\begin{equation}
 r_{\rm max}=e^5\simeq 148.41,\qquad
 e^6\simeq 403.43,\qquad
e^7\simeq 1096.63 
\end{equation}
and recalculate $\alpha_{\rm th(q)}$ for $q=0.5$, $2$, and $20$.
The results are summarized in Table~\ref{tab:conv-rmax}.
\begin{table}[ht]
\centering
\caption{Convergence of the threshold coupling $\alpha_{\rm th}(q)$ with
respect to the outer numerical boundary $r_{\rm max}$.}
\label{tab:conv-rmax}
\begin{tabular}{c|ccc|c}
\hline
$q$
& $\alpha_{\rm th}(e^5)$
& $\alpha_{\rm th}(e^6)$
& $\alpha_{\rm th}(e^7)$
& $\epsilon_{r_{\rm max}}(\%)$ \\
\hline
0.5 & $19.627554$ & $19.628877$ & $19.629013$ & $0.000695482$ \\
2   & $-0.97620131$ & $-0.97621660$ & $-0.97621863$ & $0.000207787$ \\
20  & $-0.2927210$ & $-0.2926661$ & $-0.2926703$ & $0.00143079$ \\
\hline
\end{tabular}
\end{table}
In Table 2, the relative change is defined as
\begin{equation}
 \epsilon_{r_{\rm max}}(\%)
 =
 \left|
 \frac{\alpha_{\rm th}(e^7)-\alpha_{\rm th}(e^6)}
 {\alpha_{\rm th}(e^7)}
 \right|\times 100\%.
\end{equation}
The very small values of $\epsilon_{r_{\rm max}}$ indicate that
$\alpha_{\rm th}(q)$ is well converged with respect to the location
of the outer numerical boundary.

Finally, we also vary the near-horizon cutoff,
\begin{equation}
 r_a=(1+\epsilon_h)r_+,\qquad
 \epsilon_h=10^{-2},\,10^{-3},\,10^{-4}
\end{equation}
and recompute  $\alpha_{\rm th}(q)$ for the same  $q$.
The results are shown in Table~\ref{tab:conv-horizon}.
\begin{table}[htbp]
\centering
\caption{Convergence of $\alpha_{\rm th}(q)$ with
respect to the near-horizon cutoff $\epsilon_h$.}
\label{tab:conv-horizon}
\begin{tabular}{c|ccc}
\hline
$q$
& $\alpha_{\rm th}(\epsilon_h=10^{-2})$
& $\alpha_{\rm th}(\epsilon_h=10^{-3})$
& $\alpha_{\rm th}(\epsilon_h=10^{-4})$ \\
\hline
0.5 & $19.62886963$ & $19.62887702$ & $19.62886434$ \\
2   & $-0.976216410$ & $-0.976216608$ & $-0.976217557$ \\
20  & $-0.292665965$ & $-0.292666108$ & $-0.292666204$ \\
\hline
\end{tabular}
\end{table}
The weak dependence of $\alpha_{\rm th}(q)$ on both $r_{\rm max}$ and
$\epsilon_h$ demonstrates that the threshold couplings are insensitive
to the finite outer boundary and the near-horizon cutoff, confirming the
numerical convergence of our results.

{\bfseries
To further examine the numerical convergence of the scalar
perturbation spectrum, we calculate representative purely imaginary
QNM frequencies on the scalarized branches for $q=0.5(n=0)$,
$2$, and $20$. We vary both the near-horizon cutoff $\epsilon_h$
and the outer numerical boundary $r_{\max}$. The results are
summarized in Table~4. The weak dependence of $\Omega$ on both $\epsilon_h$ and
$r_{\max}$ shows that the representative purely imaginary
QNM frequencies are numerically well converged.

\begin{table}[htbp]
\centering
\caption{Convergence of representative purely imaginary
QNM  frequencies.}
\begin{tabular}{c|ccc}
\hline
Numerical setup
& $q=0.5,\ \alpha=25$
& $q=2,\ \alpha=-0.90$
& $q=20,\ \alpha=-0.290$ \\
\hline
$\epsilon_h=10^{-3},\ r_{\max}=115 r_h$
& -0.00541980 & -0.0972926 & -0.026274974 \\

$\epsilon_h=10^{-3},\ r_{\max}=120 r_h$
& -0.00513450 & -0.0955960 & -0.026000025 \\

$\epsilon_h=10^{-3},\ r_{\max}=125 r_h$
& -0.00487555 & -0.0936358 & -0.025681493 \\

$\epsilon_h=10^{-2},\ r_{\max}=120 r_h$
&-0.00513450 &-0.0955962 & -0.026000000 \\

$\epsilon_h=10^{-1},\ r_{\max}=120 r_h$
& -0.00513449 & -0.0946809 & -0.026000002 \\
\hline
\end{tabular}
\end{table}

\par}

\section{Derivation of linearized scalar equation around scalarized EEHBH}
In this appendix, we derive the explicit form of the linearized
scalar equation around a scalarized EEHBH by considering the radial
perturbations in Eq.(\ref{p-metric}). In this case, we have to use decoupling process to find the $s$-scalar equation since the metric, NED field, and scalar are coupled. 
The  metric, vector potential, and scalar are considered as perturbations around a scalarized EEHBH background  
\begin{equation}
g_{\mu\nu}=\bar{g}_{\mu\nu}+\epsilon\delta g_{\mu\nu}, A_\mu=\bar{A}_{\mu}+\epsilon\delta A_{\mu}, \phi = \phi(r) + \epsilon \delta \tilde{\phi},
 \label{eq:metric-EM-SC}
\end{equation}
where $\epsilon$ denotes an arbitrarily small quantity.
In this case, the metric perturbation for polar-part takes the form 
\begin{align}
&\delta g_{\mu\nu} = \sum_{l,m}\int d\omega\,e^{-i\omega t}Y_{l}^{\mathfrak{m}}(\theta,\hat{\varphi}) \times
\begin{bmatrix}
N(r) e^{-2 \delta (r)}H_0(r) & 0 & 0 & 0 \\
* & \frac{H_2(r)}{N(r)} & 0 & 0 \\
* & * & r^{2} & 0 \\
* & * & * & r^{2}\sin^{2}\theta\
\end{bmatrix}.\label{eq:metricpert} &
\end{align}
where $Y_{l}^{\mathfrak{m}}(\theta,\hat{\varphi})$ are spherical harmonics. 
The vector perturbation is composed of  $\delta A_{\mu}=\delta A_{\mu}^{A}+\delta A_{\mu}^{P}$ where  $\delta A_{\mu}^{A}$ and $\delta A_{\mu}^{P}$ are given by 
\begin{flalign}
\delta A_{\mu}^{A} &= \sum_{l,m}\int d\omega\,e^{-i\omega t}  \left(0,0,-\frac{u_{4}(r)\partial_{\phi}Y_{l}^{\mathfrak{m}}}{\sin\theta},\,u_{4}(r)\sin\theta\,\partial_{\theta}Y_{l}^{\mathfrak{m}}\right), \label{eq:axialEMpert}\\
\delta A_{\mu}^{P} &= \sum_{l,m}\int d\omega\,e^{-i\omega t} \Big(u_{1}(r)Y_{l}^{\mathfrak{m}},\,u_{2}(r)Y_{l}^{\mathfrak{m}},\,u_{3}\partial_{\theta}Y_{l}^{\mathfrak{m}},u_{3}\partial_{\phi}Y_{l}^{\mathfrak{m}}\Big), \label{eq:polarEMpert}
\end{flalign}
where $u_i$ for $i=1,2,3,4$ represent static vector perturbations. 
Finally, we express  the scalar perturbation as
\begin{equation}
\delta\tilde{\phi}=\sum_{l,m}\int d\omega\,e^{-i\omega t} \frac{\tilde{\varphi}(r)}{r} Y_{l}^{\mathfrak{m}}.
\label{eq:scpert}
\end{equation}
\subsection{Linearized Einstein equation}
Substituting the perturbations into Eq.\eqref{equa1}, we obtain the linearized Einstein equation
\begin{equation}
 \delta\mathcal{E}_{\mu\nu}
 \equiv \delta G_{\mu\nu}-\delta T^{\phi}_{\mu\nu}-\delta T^{F}_{\mu\nu}.
 \label{eq:Einsteinpert}
\end{equation}
For the case with $s$-mode (\(l=0\) and  \(\mathfrak{m}=0\)), the electromagnetic contribution to the perturbed energy-momentum tensor ($\delta T^{F}_{\mu\nu}$) is of second order \(\epsilon^{2}\). The quantities $\delta F_{\mu\rho}F_{\nu}{}^{\rho}$ and $\delta \mathcal{F}$ present in the perturbed electromagnetic energy-momentum tensor $\delta T^{F}_{\mu\nu}$ and the perturbed scalar Eq.\eqref{s-equa} have their lowest-order terms of second order $\epsilon^2$.
\begin{eqnarray}
&&\delta F_{\mu\rho}F_{\nu}~^\rho=\epsilon^2\begin{bmatrix}
\frac{N(r) (i\omega u_2+u_1')^2}{4\pi} & 0 & 0 & 0 \\
0 & \frac{e^{2\delta} (\omega u_2-i u_1')^2}{4\pi N} & 0 & 0 \\
0 & 0 & 0 & 0 \\
0 & 0 & 0 & 0
\end{bmatrix}\\
&&\delta \mathcal{F}=\epsilon^2\frac{e^{2\delta} (\omega u_2-i u_1')^2}{2\pi}
\label{deltaF}.
\end{eqnarray}
Thus, these second‑order terms can be neglected when considering the $s$-mode linearized theory.  Hence, the electromagnetic perturbation is decoupled from the linear Einstein equation and the linear scalar equation.

After extracting the time factor $e^{-i\omega t}$, we  redefine its four components as 
\begin{align}
 \delta\mathcal{E}_{tt}&=
 \frac{e^{-2\delta-\alpha\phi^2}N}{2\sqrt{\pi}\,r^9}
 \mathcal{G}_{tt},
 &
 \delta\mathcal{E}_{tr}&=-\frac{i\omega}{2\sqrt{\pi}\,r}
 \mathcal{G}_{tr},
 \label{eq:normalization-1}\\
 \delta\mathcal{E}_{rr}&=-
 \frac{e^{-\alpha\phi^2}}{2\sqrt{\pi}\,r^9N}
 \mathcal{G}_{rr},
 &
 \delta\mathcal{E}_{\theta\theta}&=
 \frac{1}{8\sqrt{\pi}\,r^7}\mathcal{G}_{\theta\theta}.
 \label{eq:normalization-2}
\end{align}
Then, four components of linearized Einstein equation can be managed as follows. 
The $tt$-equation is given by 
\begin{flalign}
 0=\mathcal{G}_{tt}=&
 2\alpha q^2\left(r^4-2\mu q^2\right)\phi\tilde{\varphi}
 \nonumber\\
 &+e^{\alpha\phi^2}r^7\Bigl\{
 rNH_2'+2N\tilde{\varphi}\phi'-2rN\phi'\tilde{\varphi}'
 \nonumber\\
 &+H_0\left[-1+N+rN'+r^2N(\phi')^2\right]
 +H_2\left[N+rN'+r^2N(\phi')^2\right]
 \Bigr\}.
 \label{eq:Gtt}
\end{flalign}
The $tr$-equation provides  a simple constraint
\begin{flalign}
 0=\mathcal{G}_{tr}=H_2-2\tilde{\varphi}\phi'.
 \label{eq:Gtr}
\end{flalign}
The $rr$-equation takes the form 
\begin{flalign}
 0=\mathcal{G}_{rr}={}&
 e^{\alpha\phi^2}r^7
 \left\{
 H_2+N\left[rH_0'-2\phi'\left(\tilde{\varphi}-r\tilde{\varphi}'\right)\right]
 \right\}
 \nonumber\\
 &+2\alpha q^2\left(r^4-2\mu q^2\right)\phi\tilde{\varphi}.
 \label{eq:Grr}
\end{flalign}
Also,  the $\theta\theta$-component has a complicated form 
\begin{align}
 0=\mathcal{G}_{\theta\theta}={}&
 8\alpha q^2 e^{-\alpha\phi^2}
 \left(r^4-6\mu q^2\right)\phi\tilde{\varphi}
 -r^9N'\left(3H_0'+H_2'\right)
 \nonumber\\
 &+2r^7N\Bigl[
 rH_2'\left(r\delta'-1\right)
 +rH_0'\left(2r\delta'-1\right)
 -4\tilde{\varphi}\phi'+4r\phi'\tilde{\varphi}'-r^2H_0''
 \Bigr]
 \nonumber\\
 &+\frac{2r^8H_2}{N}\Bigl\{
 N\left[N'\left(3r\delta'-2\right)
 -rN''\right]+r\omega^2 e^{2\delta}
 \nonumber\\
 &-2N^2\left[r(\delta')^2
 -\delta'+r\left((\phi')^2-\delta''\right)\right]
 \Bigr\}.
 \label{eq:Gtheta}
\end{align}
Finally, we comment that the $\varphi\varphi$-equation is a redundant one because  of 
$\delta\mathcal{E}_{\varphi\varphi}=\sin^2\theta\,\delta\mathcal{E}_{\theta\theta}$.

\subsection{Elimination of the metric perturbations from the scalar equation}
Initially, the perturbed equation is coupled to the metric perturbations $H_0$ and $H_2$ as 
\begin{align}
 0=\mathcal{K}={}&
 \frac{N}{r}\tilde{\varphi}''
 +\frac{N'-N\delta'}{r}\tilde{\varphi}'
 -\frac{N}{2}\phi'\left(H_0'+H_2'\right)
 \nonumber\\
 &+\left\{
 \frac{N}{r}\left[\left(r\delta'-2\right)\phi'-r\phi''\right]
 -N'\phi'
 \right\} H_2
 \nonumber\\
 &+\left\{
 \frac{\omega^2e^{2\delta}}{rN}
 +\frac{N\delta'-N'}{r^2}
 +\frac{\alpha q^2\left(r^4-2\mu q^2\right)
 e^{-\alpha\phi^2}\left(1-2\alpha\phi^2\right)}{r^9}
 \right\}\tilde{\varphi}.
 \label{eq:KG-before-constraints}
\end{align}
From Eq.\eqref{eq:Gtr}, one finds  one relation and its first derivative 
\begin{equation}
 H_2=2\phi'\tilde{\varphi},
 \qquad
 H_2'=2\phi'\tilde{\varphi}'+2\phi''\tilde{\varphi}.
 \label{eq:H1-solution}
\end{equation}
Substituting Eq.\eqref{eq:H1-solution} into Eq.~\eqref{eq:Grr} and solving for $H'_0$,
we obtain the following relation between $H_0$ and $\tilde{\varphi}$ as 
\begin{align}
 H_0'={}&-2\phi'\tilde{\varphi}'
 +\left[
 \frac{2(N-1)\phi'}{rN}
 -\frac{2\alpha q^2\left(r^4-2\mu q^2\right)
 e^{-\alpha\phi^2}\phi}{r^8N}
 \right]\tilde{\varphi}.
 \label{eq:H0prime-solution}
\end{align}
Eq.~\eqref{eq:KG-before-constraints} becomes a decoupled one, thanks to Eqs.\eqref{eq:H1-solution} and \eqref{eq:H0prime-solution}.  After normalization of  a coefficient in front of $\tilde{\varphi}''$, one obtains a decoupled scalar equation 
\begin{equation}
 \tilde{\varphi}''+\left(\frac{N'}{N}-\delta'\right)\tilde{\varphi}'
 +\left[\frac{\omega^2 e^{2\delta}}{N^2}
 +\mathcal{U}_0(r)\right]\tilde{\varphi}=0,
 \label{eq:reduced-radial}
\end{equation}
where $\mathcal{U}_0(r) $ is expressed in terms of scalarized EEHBH variables as 
\begin{align}
 \mathcal{U}_0={}&
 \delta'\left(\frac{1}{r}+2r(\phi')^2\right)
 -\phi'\left(5\phi'+3r\phi''\right)
 \nonumber\\
 &+\frac{1}{r^8N}\Bigl\{
 \alpha q^2\left(r^4-2\mu q^2\right)e^{-\alpha\phi^2}
 \left(1-2\alpha\phi^2+r\phi\phi'\right)
 \nonumber\\
 &
 +r^8(\phi')^2-r^7N'\left[1+2r^2(\phi')^2\right]
 \Bigr\}.
 \label{eq:U0}
\end{align}

\subsection{Schr\"odinger-type equation with potential}
We use the background Einstein equations to simplify the perturbed scalar  equation as a final step. 
From  $\mathcal{E}_{tt}=0$ and $\mathcal{E}_{rr}=0$,  two relations are found as 
\begin{align}
 \delta'={}&-r(\phi')^2,
 \label{eq:bg-delta}\\
 N'={}&\frac{e^{-\alpha  \phi^2} \left(2 \mu  q^4-q^2 r^4+r^6 e^{\alpha  \phi^2}\right)}{r^7}-\frac{N \left(r^2 \phi'^2+1\right)}{r}.
 \label{eq:bg-N}
\end{align}
From the background scalar equation, we obtain
\begin{align}
\phi''={}&
 \frac{
 r^7\phi'\left[N\left(r\delta'-2\right)-rN'\right]
 +\alpha q^2\phi\left(2\mu q^2-r^4\right)e^{-\alpha\phi^2}
 }{r^8N}.
 \label{eq:bg-phi}
\end{align}

Making use of the above relations, we can simplify $\mathcal{U}_0(r)$ in Eq.\eqref{eq:U0}.
Now, let us  introduce  the tortoise coordinate defined by $\frac{d r_*}{d r}=\frac{e^{\delta}}{N}$ and transform Eq.\eqref{eq:reduced-radial} into the tortoise coordinate. Under this transformation, the coefficient of the \(\tilde{\varphi}'(r_*)\) becomes $\frac{d^2 r_*}{d r^2}+\left(\frac{N'}{N}-\delta'\right)\frac{d r_*}{d r}=0$, 
leading to cancellation of the $\tilde{\varphi}'$-term. 

Finally, upon normalizing the resulting equation, we recover the Schr\"odinger-type  Eq.(\ref{tiS-eq}) with potential $V_{\rm sEEH}(r,q,\alpha)$.

\end{document}